\documentclass[reprint,twocolumn]{revtex4}
\usepackage{amsfonts}
\usepackage{amsmath}
\usepackage{amssymb}
\usepackage{charter}
\usepackage{subfigure}
\usepackage{graphicx}
\usepackage{float}

\begin{document}

\title{Quantum-improved phase estimation with a displacement-assisted
SU(1,1) interferometer}
\author{Wei Ye$^{1}$}
\author{Shoukang Chang$^{2}$}
\author{Shaoyan Gao$^{2}$}
\author{Huan Zhang$^{3}$}
\thanks{Corresponding author. zhangh739@mail2.sysu.edu.cn}
\author{Ying Xia$^{3}$}
\thanks{Corresponding author. xiay78@mail2.sysu.edu.cn}
\author{Xuan Rao$^{2}$}
\thanks{Corresponding author. raoxuancom@163.com}
\affiliation{$^{{\small 1}}$\textit{School of Information Engineering, Nanchang Hangkong
University, Nanchang 330063, China}\\
$^{{\small 2}}$\textit{MOE Key Laboratory for Nonequilibrium Synthesis and
Modulation of Condensed Matter, Shaanxi Province Key Laboratory of Quantum
Information and Quantum Optoelectronic Devices, School of Physics, Xi'an
Jiaotong University, 710049, People's Republic of China}\\
$^{{\small 3}}${\small \ }\textit{School of Physics, Sun Yat-sen University,
Guangzhou 510275, China}}

\begin{abstract}
By performing two local displacement operations (LDOs) inside an SU(1,1)
interferometer, called as the displacement-assisted SU(1,1) [DSU(1,1)], both
the phase sensitivity based on homodyne detection and quantum Fisher
information (QFI) with and without photon losses are investigated in this
paper. In this DSU(1,1) interferometer, we focus our attention on the extent
to which the introduced LDO affects the phase sensitivity and the QFI, even
in the realistic scenario. Our analyses show that the estimation performance
of DSU(1,1) interferometer is always better than that of SU(1,1)
interferometer without the LDO, especially the phase precision of the former
in the ideal scenario gradually approaching to the Heisenberg limit via the
increase of the LDO strength. More significantly, different from the latter,
the robustness of the former can be enhanced markedly by regulating and
controlling the LDO. Our findings would open an useful view for
quantum-improved phase estimation of optical interferometers.

\end{abstract}

\maketitle

\section{Introduction}

Quantum metrology is an excellent candidate of parameter estimation theory
to serve as the high-precision requirement of various quantum information
tasks \cite{1,2,3,4,5,6}, such as quantum sensor \cite{1,2} and quantum
imaging \cite{3,4}. Thus, how to achieve the higher precision of quantum
metrology has become a general consensus among scientists. To this end, the
optical interferometers, e.g., Mach-Zehnder interferometer (MZI) \cite%
{7,8,9,10,11,12,13} and SU(1,1) interferometer \cite{14,15,16,17,18,19,20,21}%
, are often used for understanding the subtle phase variations thoroughly.
Generally, in optical-interferometer systems, it is possible to obtain the
higher precision of phase estimation using three probe strategies:
generation, modification and readout \cite{22}.
\begin{figure*}[tbp]
\centering \includegraphics[width=1.2\columnwidth]{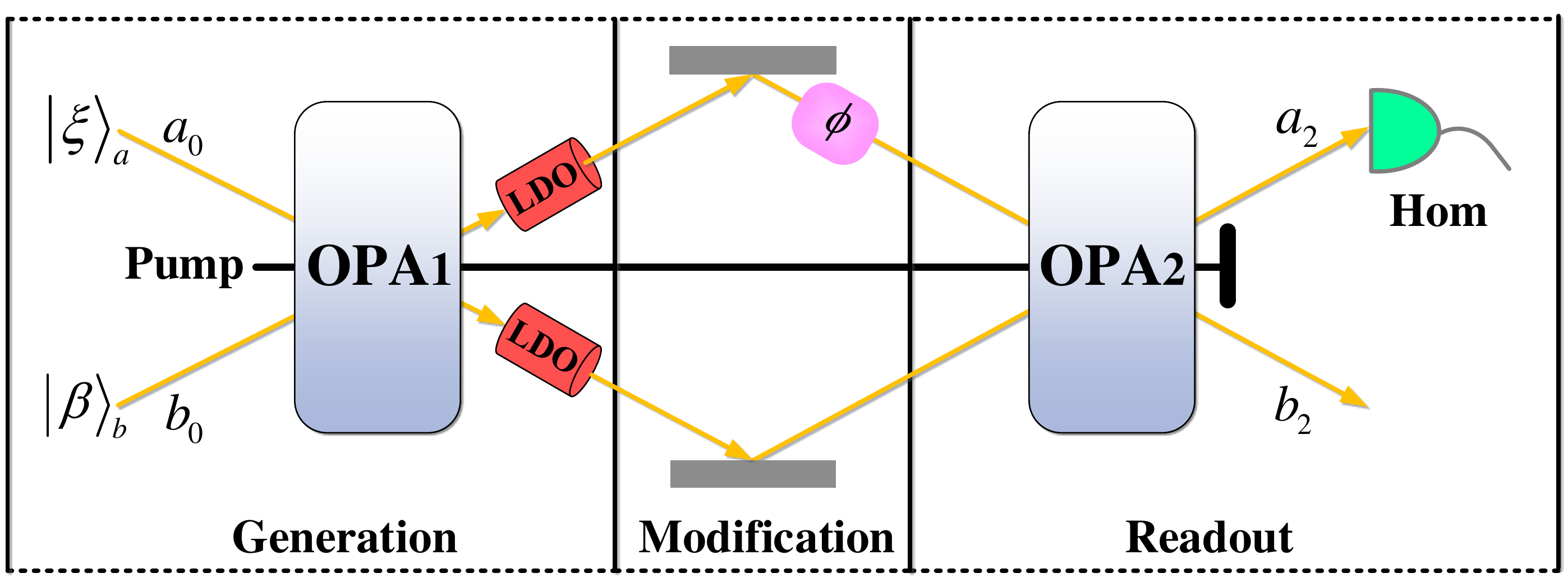}
\caption{{}(Color online) Schematic diagram of the DSU(1,1) interferometer
together with homodyne detection, in which a squeezed vacuum state $%
\left
\vert \protect \xi \right \rangle _{a}$ and a coherent state $%
\left
\vert \protect \beta \right \rangle _{b}$ are respectively used as the
inputs of DSU(1,1) interferometer in paths $a$ and $b$. OPA$_{1}$ and OPA$%
_{2}$: the first and second optical parametric amplifier. LDO is a local
displacement operation. $\protect \phi $ is a phase shift to be measured.
Hom: an homodyne detection. $a_{0}(b_{0})$ and $a_{2}(b_{2})$: the input and
output operators of DSU(1,1) interferometer, respectively. }
\end{figure*}
In the probe generation stage, nonclassical quantum resources as the inputs
of the MZI have been proven to more effectively enhance the precise
measurement than its classical counterpart \cite{23,24,25,26,27}. In
particular, when using the NOON states \cite{25,28}, the two-mode squeezed
vacuum states (TMSVS) \cite{26} and the twin Fock states \cite{27}, the
standard quantum limit (SQL) \cite{13} that is not exceeded by only
exploiting the classical resources can be easily beaten, even infinitely
reaching at the famed Heisenberg limit (HL) \cite{18,26}. These states,
however, are extremely sensitive to noisy environments \cite{29}, so that
non-Gaussian resources \cite{12,30,31,32,33} as an alternative that can be
produced by taking advantage of non-Gaussian operations on an arbitrary
initial state play an important role in improving the estimation performance
of the MZI, even in the presence of noisy scenarios \cite{12,33}. Apart from
the generation stage, many efforts have devoted to conceiving the probe
modification by replacing the conventional beam splitters in the
conventional MZI with the optical parametric amplifiers (OPAs) \cite%
{14,15,16,17,18,21,22,34}, which is also called as SU(1,1) interferometer
proposed first by Yurke \cite{16}. In this SU(1,1) interferometer with two
OPAs, the first OPA (denoted as OPA$_{1}$) is used not only to obtain the
entangled resources but also to eliminate amplified noise; while the usage
of the second OPA (denoted as OPA$_{2}$) can result in the signal
enhancement \cite{21,22}, which paves a feasible way to achieve the higher
precision of phase estimation. Taking advantage of these features, an
SU(1,1) interferometer scheme with the phase shift induced by a kerr medium
was suggested by Chang \cite{22}, pointing out that the significant
improvement of both the phase sensitivity and quantum Fisher information can
be achieved even in the presence of photon losses. In addition, the
noiseless quantum amplification of parameter-dependent processes was used to
SU(1,1) interferometer, indicating how this process results in the HL \cite%
{35}. More interestingly, by using the non-Gaussian operations inside the
SU(1,1) interferometer, both the phase sensitivity and the robustness of
this interferometer system against the photon losses can be further enhanced
\cite{36}. From works \cite{12,23,33,36}, we also notice that the usage of
non-Gaussian operations can significantly improve the estimation performance
of the optical interferometers, but at the expense of the high cost of
implementing these operations. \  \  \  \  \  \  \

To solve the above problem, the local operations containing the local
squeezing operation (LSO) \cite{37,38,39} and the local displacement
operation (LDO) \cite{40} are one of the most promising choices. In
particular, J. Sahota and D. F. V. James suggested a quantum-enhanced phase
estimation scheme by applying the LSO into the MZI \cite{39}. However, it
should be mentioned that the LSO plays a key role in quantum metrology \cite%
{39}, quantum key distribution \cite{38} and entanglement distillation \cite%
{37}, but the degree of the LSO is not infinite, e.g., its maximum
attainable degree for the TMSVS about $1.19$ ($10.7$ dB) \cite{41}. For this
reason, here we suggest a quantum-improved phase estimation of the SU(1,1)
interferometer based on the LDO, which can be called as the
displacement-assisted SU(1,1) [DSU(1,1)] interferometer. Under the framework
of this DSU(1,1)] interferometer, we not only derive its explicit forms of
both the quantum Fisher information (QFI) and the phase sensitivity based on
homodyne detection, but also consider the effects of photon losses on its
estimation performance. Our analyses manifest that the increase of the LDO
strength is conducive to the improvement of both the QFI and the phase
sensitivity, even in the presence of photon losses. In particular, this
increasing LDO can narrow the gap for the phase sensitivity between with and
without photon losses. This implies that the usage of the sufficiently large
LDO can make the SU(1,1) interferometer systems more robust against photon
losses.

The remainder of this paper is arranged as follows. In section II, we first
describe the theoretical model of DSU(1,1) interferometer, and then give the
relationship between the output and input operators for this interferometer.
In sections III, for the ideal scenario, we analyze and discuss both the QFI
and the phase sensitivity based on homodyne detection in DSU(1,1)
interferometer, before making a comparison about phase sensitivities
containing the SQL, the HL and the DSU(1,1) interferometer scheme in section
VI. Subsequently, we also consider the effects of photon losses on both the
QFI and the phase sensitivity of DSU(1,1) interferometer in section V.
Finally, our main conclusions are drawn in the last section.
\begin{figure*}[tbp]
\label{Fig2} \centering \includegraphics[width=0.72\columnwidth]{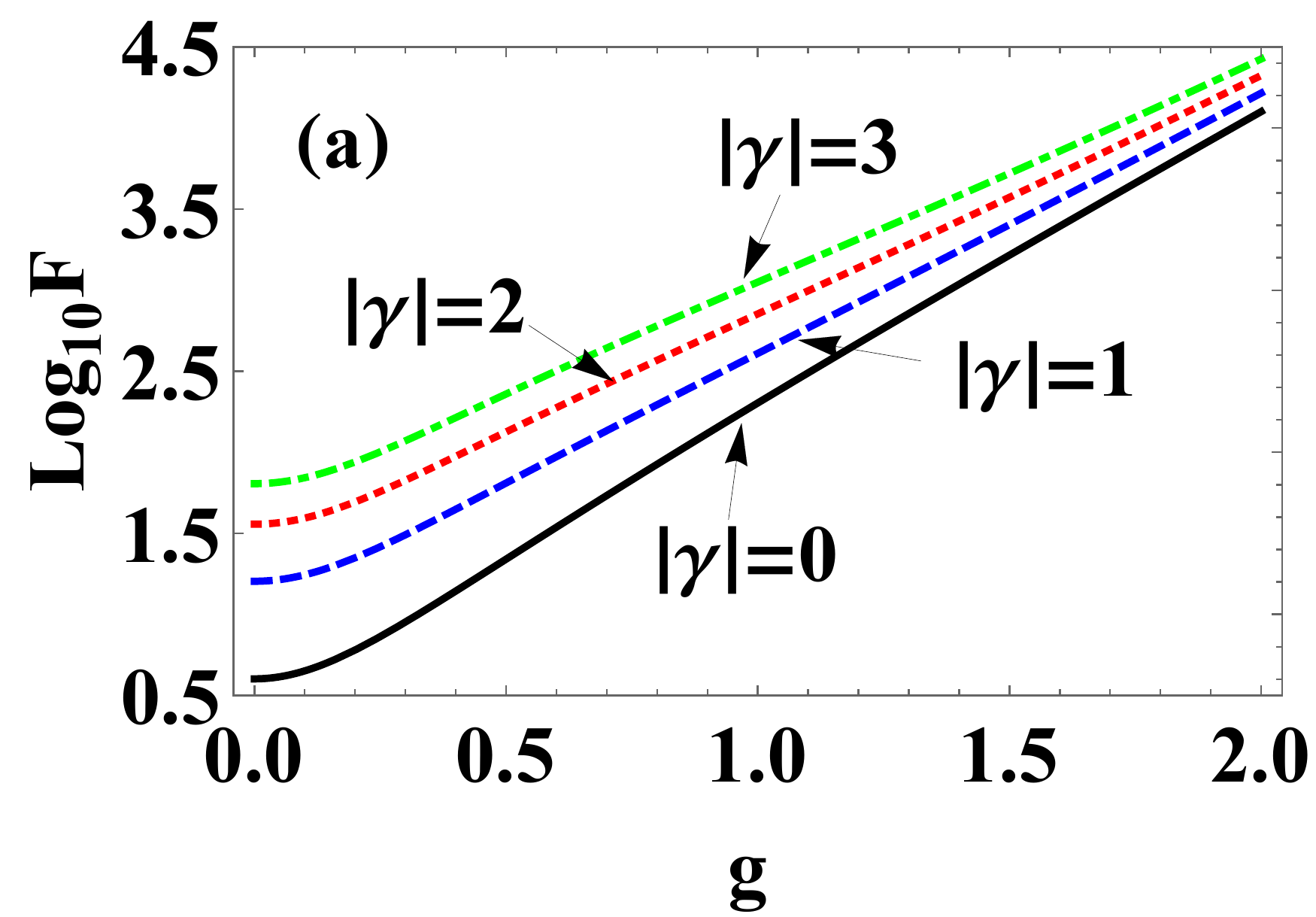}%
\includegraphics[width=0.72\columnwidth]{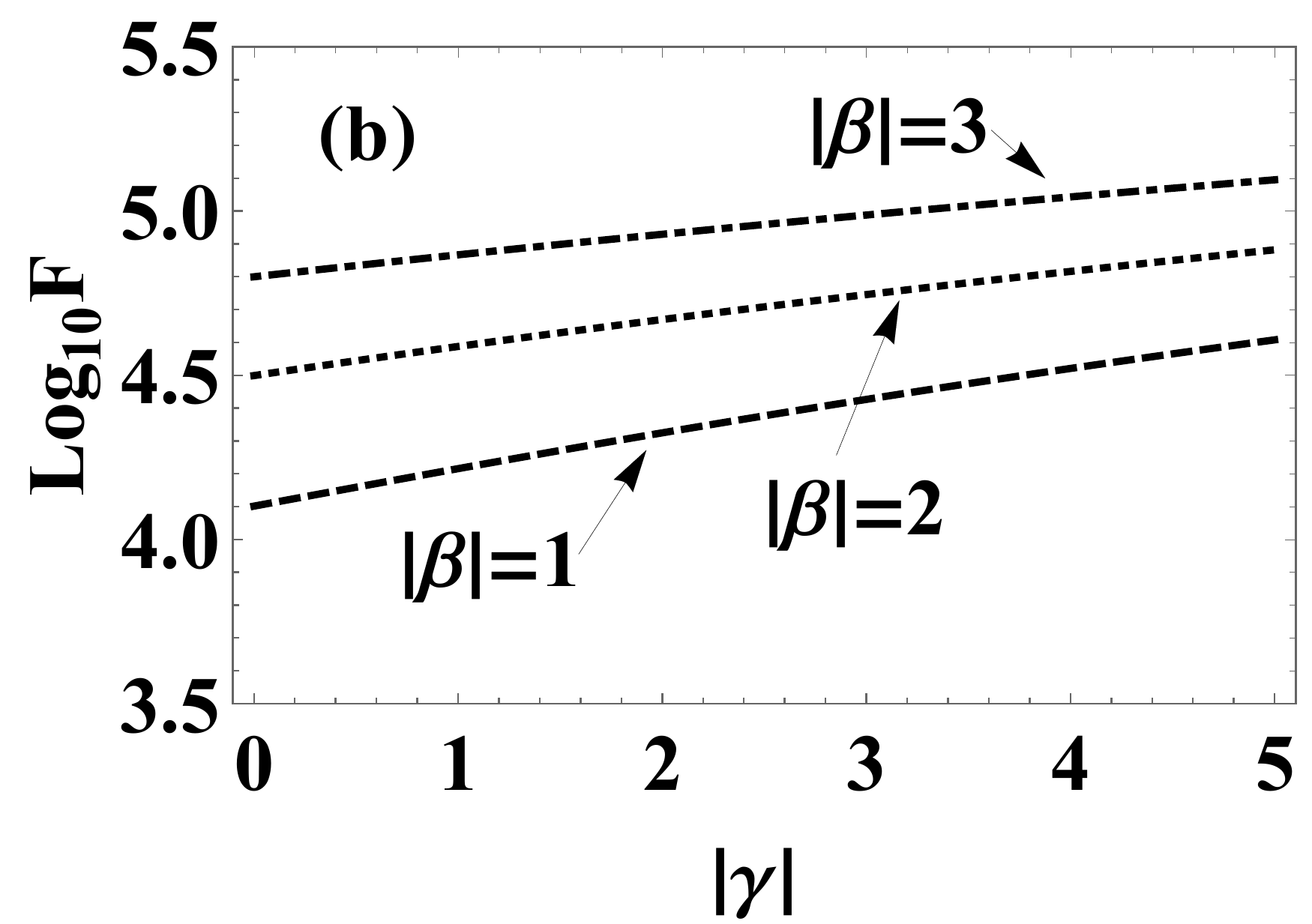} %
\includegraphics[width=0.72\columnwidth]{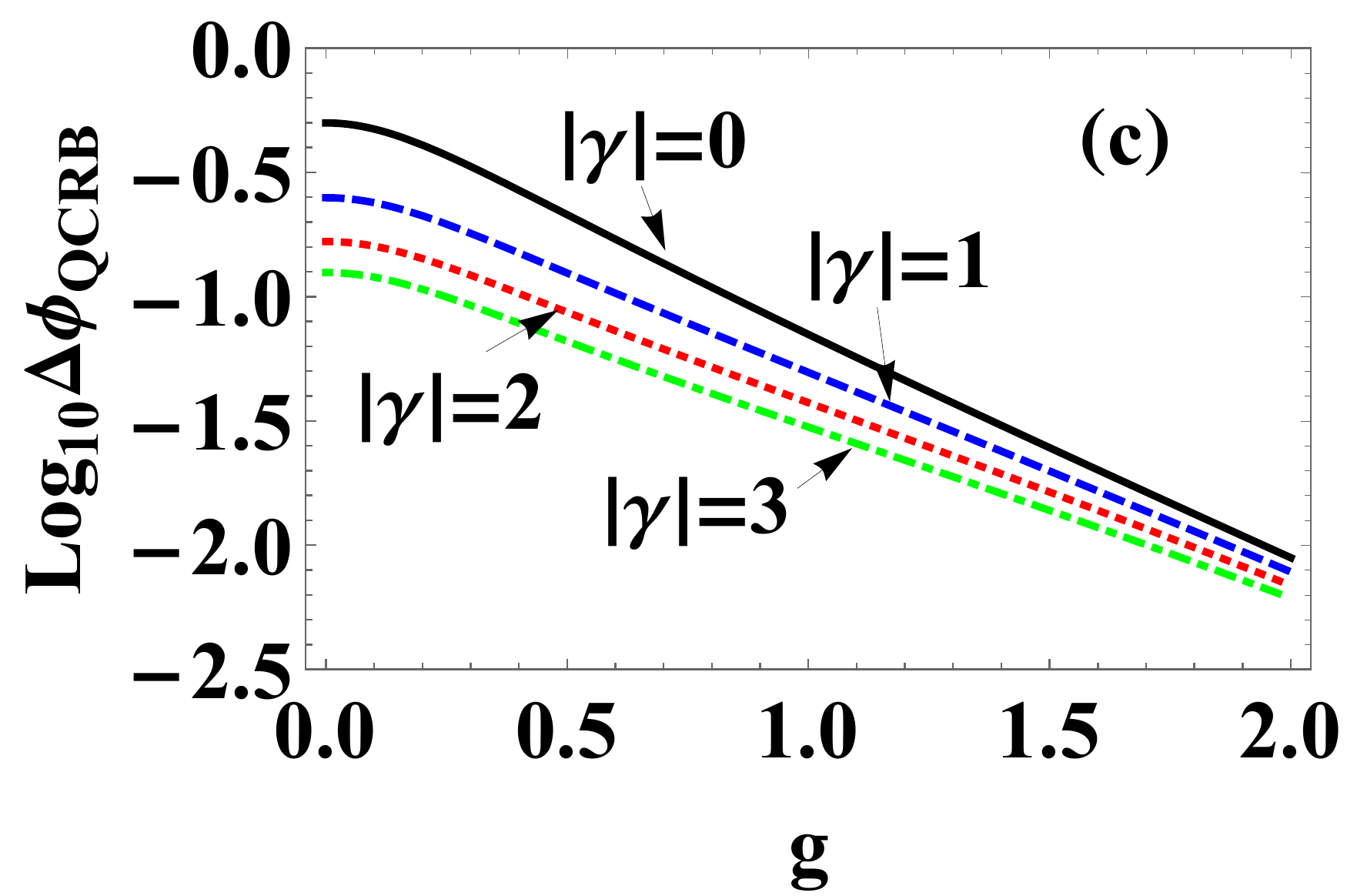} %
\includegraphics[width=0.72\columnwidth]{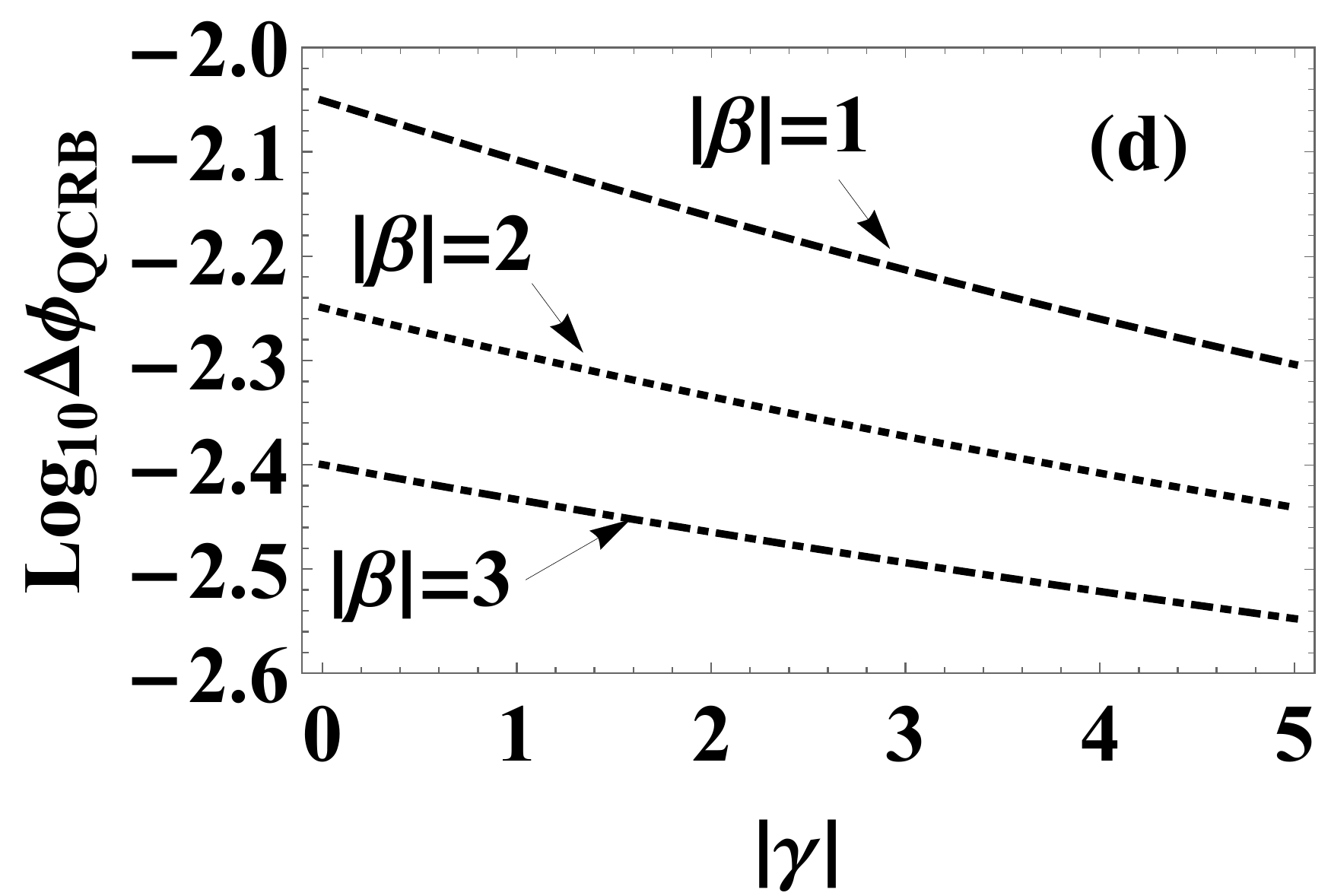}
\caption{{}(Color online) Both (a)-(b) the QFI $\log _{10}F$ and the QCRB $%
\log _{10}\Delta \protect \phi _{QCRB}$ as a function of $g$ at (a) and (c)
with fixed values $\left \vert \protect \beta \right \vert =r=1$ for
different $\left \vert \protect \gamma \right \vert =0,1,2,3,$ and of $%
\left
\vert \protect \gamma \right \vert $ at (b) and (d) with fixed values $%
g=2,r=1 $ for different $\left \vert \protect \beta \right \vert =1,2,3.$
Other parameters are as following: $\protect \phi =\protect \theta _{\protect%
\xi }=0$ and $\protect \theta _{\protect \beta }=\protect \theta _{\protect%
\gamma }=\protect \pi /2$.}
\end{figure*}

\section{ The DSU(1,1) interferometer and its relationship between the
output and input operators}

Now, let us\ begin with introducing the theoretical model of DSU(1,1)
interferometer, whose structure is comprised of two OPAs, two LDOs and a
linear phase shift, as depicted in Fig. 1. For simplicity, here we only
consider both a squeezed vacuum state $\left \vert \xi \right \rangle _{a}=%
\hat{S}(\xi )\left \vert 0\right \rangle _{a}$ with the squeezing operator $%
\hat{S}(\xi )=\exp [(\xi ^{\ast }\hat{a}^{2}-\xi \hat{a}^{\dagger 2})/2]$ ($%
\xi =re^{i\theta _{\xi }}$) on the vacuum state $\left \vert 0\right \rangle
_{a}$ and a coherent state $\left \vert \beta \right \rangle _{b}$ with $%
\beta =\left \vert \beta \right \vert e^{i\theta _{\beta }}$ as the inputs
of DSU(1,1) interferometer in paths $a$ and $b$, respectively. After these
input states pass through the OPA$_{1}$, paths $a$ and $b$ respectively
experience the same LDO process, denoted as $\hat{D}_{a}(\gamma )=e^{\gamma
\hat{a}^{\dagger }-\gamma ^{\ast }\hat{a}}$ and $\hat{D}_{b}(\gamma
)=e^{\gamma \hat{b}^{\dagger }-\gamma ^{\ast }\hat{b}}$ with $\gamma
=\left
\vert \gamma \right \vert e^{i\theta _{\gamma }}$, so that the probe
state $\left \vert \psi _{\gamma }\right \rangle $ can be achieved. Then, we
also assume that path $a$ serves as the reference path, while path $b$
undergoes a linear phase shifter for producing a phase shift $\phi $ to be
estimated. Finally, after paths $a$ and $b$ recombine in the OPA$_{2}$, we
can extract the phase information about the value of $\phi $ by implementing
the homodyne detection in path $a$. Indeed, the relationship between the
output and input operators for DSU(1,1) interferometer can be given by%
\begin{eqnarray}
\hat{a}_{2} &=&W_{1}+Y\hat{a}_{0}-Z\hat{b}_{0}^{\dagger },  \notag \\
\hat{b}_{2} &=&W_{2}+e^{i\phi }(Y\hat{b}_{0}-Z\hat{a}_{0}^{\dagger }),
\label{1}
\end{eqnarray}%
where $W_{1}$ and $W_{2}$ are caused by the LDO process, and
\begin{eqnarray}
Y &=&\cosh g_{1}\cosh g_{2}+e^{i(\theta _{2}-\theta _{1}-\phi )}\sinh
g_{1}\sinh g_{2},  \notag \\
Z &=&e^{i\theta _{1}}\sinh g_{1}\cosh g_{2}+e^{i(\theta _{2}-\phi )}\cosh
g_{1}\sinh g_{2},  \notag \\
W_{1} &=&\gamma \cosh g_{2}-\gamma ^{\ast }e^{i(\theta _{2}-\phi )}\sinh
g_{2},  \notag \\
W_{2} &=&\gamma e^{i\phi }\cosh g_{2}-\gamma ^{\ast }e^{i\theta _{2}}\sinh
g_{2},  \label{2}
\end{eqnarray}%
with $g_{1}(g_{2})$ and $\theta _{1}(\theta _{2})$ respectively representing
the gain factor and the phase shift in the OPA$_{1}$ (OPA$_{2}$). According
to Eq. (\ref{1}), one can further derive the explicit form of phase
sensitivity, which is a prerequisite for our analysis and discussion about
the estimation performance of DSU(1,1) interferometer in the following
sections.

\section{The QFI and phase sensitivity of \textbf{DSU(1,1) interferometer}
in an ideal scenario}

So far, we have described the schematic of DSU(1,1) interferometer in
detail. In this section, we shall present and analyze the estimation
performance of DSU(1,1) interferometer from the perspective of both quantum
Fisher information and phase sensitivity in an ideal scenario. Moreover, for
the sake of discussion, in the following sections, we also assume that the
DSU(1,1) interferometer is in the balanced case, i.e., $\theta _{2}-\theta
_{1}=\pi $ and $g_{1}=g_{2}=g$ (set $\theta _{1}=0$ and $\theta _{2}=\pi $
for simplicity).

\subsection{The QFI}

To directly assess the estimation performance of a unknown phase parameter
without any detection strategies, it is an enormous success for utilizing
the QFI of the probe state since the quantum Cram\'{e}r-Rao bound (QCRB) $%
\Delta \phi _{QCRB}$ representing the ultimate precision is in inverse
proportion to the QFI (denoted as $F$). In addition, the increased value of
the QFI indicates that the estimation precision becomes more excellent. In
this context, the QFI for an arbitrary pure state in the ideal scenario can
be expressed as \cite{12,22,32,42}

\begin{equation}
F=4[\left \langle \psi _{\phi }^{\prime }|\psi _{\phi }^{\prime }\right
\rangle -|\left \langle \psi _{\phi }^{\prime }|\psi _{\phi }\right \rangle
|^{2}],  \label{3}
\end{equation}%
where $\left \vert \psi _{\phi }\right \rangle =e^{i\phi \hat{b}^{\dagger }%
\hat{b}}\left \vert \psi _{\gamma }\right \rangle $ is the state vector
prior to the OPA$_{2}$ and $\left \vert \psi _{\phi }^{\prime
}\right
\rangle =\partial \left \vert \psi _{\phi }\right \rangle /\partial
\phi $. Thus, if the probe state is obtained, Eq. (\ref{3}) can be rewritten
as

\begin{equation}
F=4[\left \langle \psi _{\gamma }\right \vert \hat{n}^{2}\left \vert \psi
_{\gamma }\right \rangle -\left \langle \psi _{\gamma }\right \vert \hat{n}%
\left \vert \psi _{\gamma }\right \rangle ^{2}],  \label{4}
\end{equation}%
where $\hat{n}=\hat{b}^{\dagger }\hat{b}$ is the photon number operator of
path $b$. As a consequence, based on Eq. (\ref{4}), when inputting the state
$\left \vert \psi _{in}\right \rangle =\left \vert \xi \right \rangle
_{a}\otimes \left \vert \beta \right \rangle _{b}$, one can obtain the
explicit form of the QFI of DSU(1,1) interferometer (see Appendix A for more
details), i.e.,
\begin{figure*}[tbp]
\label{Fig3} \centering \includegraphics[width=0.72\columnwidth]{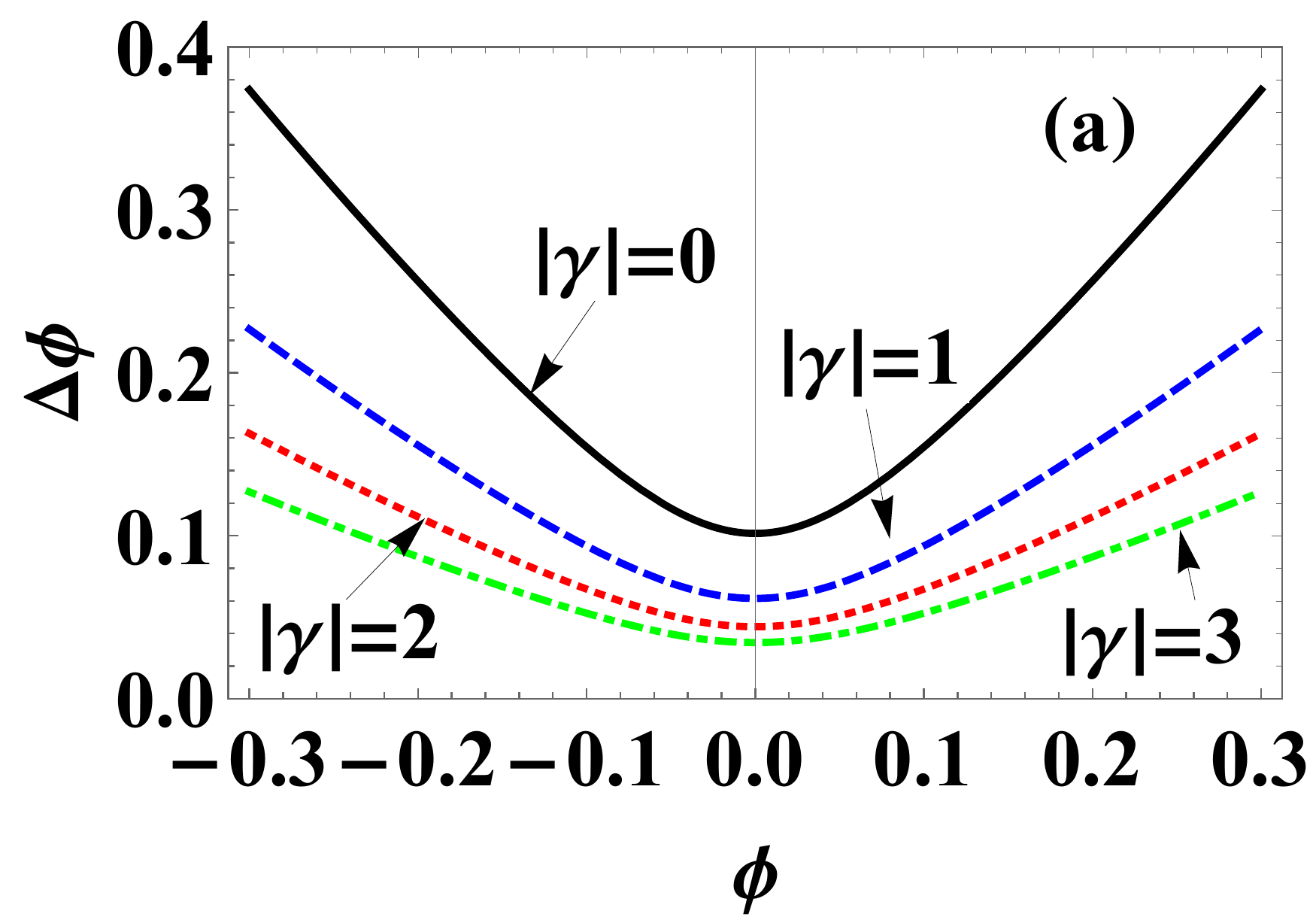}%
\includegraphics[width=0.72\columnwidth]{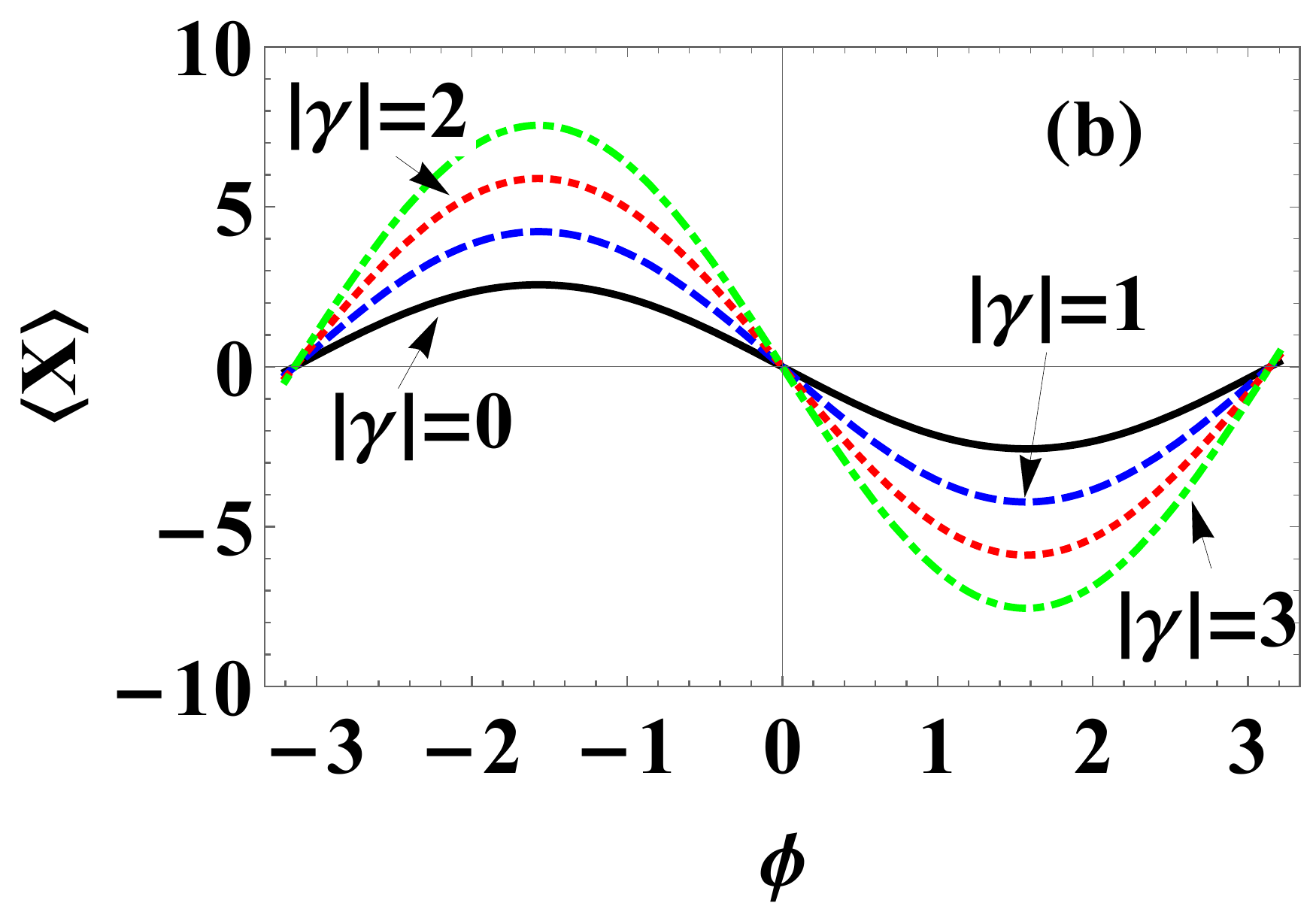}
\caption{(Color online) (a) Phase sensitivity with homodyne detection and
(b) output signal changing with $\protect \phi $ for different $\left \vert
\protect \gamma \right \vert =0,1,2,3.$ Other parameters are as following: $%
g=r=|\protect \beta |=1,$ $\protect \theta _{\protect \xi }=0$ and $\protect%
\theta _{\protect \beta }=\protect \theta _{\protect \gamma }=\protect \pi /2$.}
\end{figure*}
\begin{equation}
F=4(\Gamma _{2}+\Gamma _{1}-\Gamma _{1}^{2}),  \label{5}
\end{equation}%
where $\Gamma _{m}$ $(m=1,2)$ are the average value of operators $\hat{b}%
^{\dagger m}\hat{b}^{m}$ with respect to the probe state. By using Eq. (\ref%
{5}), one also can obtain the QCRB providing the ultimate phase precision of
DSU(1,1) interferometer regardless of detection schemes \cite{43,44}, i.e.,
\begin{equation}
\Delta \phi _{QCRB}=\frac{1}{\sqrt{\nu F}},  \label{6}
\end{equation}%
with the number of trials $\nu $ (for simplicity, set $\nu =1$).\ From Eq. (%
\ref{6}), it is obvious that, the larger the value of $F,$ the smaller the $%
\Delta \phi _{QCRB}$, which implies the attainability of the higher phase
sensitivity. In order to see this point, Fig. 2 shows both the QFI and the
QCRB changing with the gain factor $g$ and the LDO strength $\left \vert
\gamma \right \vert $. As we can see from Fig. 2(a), compared to SU(1,1)
interferometer without the LDO (the black solid line), with the increase of $%
\left \vert \gamma \right \vert =1,2,3$, the QFI can be increased well,
which would straightway lead to the decrease of the QCRB [see Fig. 2(c)],
meaning the enhanced phase sensitivity. These improvement effects of both
the QFI and the QCRB, however, can decrease with the increase of $g$. To
show the performance of both the QFI and the QCRB in the sufficiently large
case of $g $, e.g., $g=2,$ as respectively depicted in Figs. 2(b) and 2(d),
for several different $\left \vert \beta \right \vert =1,2,3$, we thus find
that both the QFI and the QCRB are still improved by increasing the value of
$\left \vert \gamma \right \vert $. These results fully indicate that the
performance of the DSU(1,1) interferometer for the ideal scenario performs
better than that of the SU(1,1) interferometer without the LDO in terms of
the QFI and the QCRB.

\subsection{The phase sensitivity via the homodyne detection}

In this subsection, we mainly focus on analyzing the phase sensitivity of
DSU(1,1) interferometer. For this purpose, it is unavoidable to choose a
specific detection scheme, such as homodyne detection \cite{22,34,45},
intensity detection \cite{15,19,46}, and parity detection \cite{12,28,25,47}%
, for reading the phase information. In addition, compared with both
intensity and parity detections, the homodyne detection is computationally
convenient and compatible with existing experimental technology, thereby
playing potential applications in quantum communication \cite{48,49,50,51,52}%
. For this reason, the phase parameter $\phi $ can be estimated by
exploiting the homodyne detection (see Fig. 1), whose detected variable can
be treated as the amplitude quadrature $\hat{X}$, i.e.,
\begin{equation}
\hat{X}=\frac{\hat{a}_{2}+\hat{a}_{2}^{\dagger }}{\sqrt{2}}.  \label{7}
\end{equation}%
Using Eq. (\ref{7}) and the error propagation formula, the phase sensitivity
of DSU(1,1) interferometer can be thus given by%
\begin{equation}
\Delta \phi =\frac{\sqrt{\Delta ^{2}\hat{X}}}{\left \vert \partial \left
\langle \hat{X}\right \rangle /\partial \phi \right \vert },  \label{8}
\end{equation}%
with $\Delta ^{2}\hat{X}=\left \langle \hat{X}^{2}\right \rangle
-\left
\langle \hat{X}\right \rangle ^{2}$. It is clearly seen from Eq. (%
\ref{8}) that, for an arbitrary value of $\phi $, the corresponding phase
sensitivity can be analytically derived, which can refer to Appendix B for
more details.

To find the optimal point $\phi $ corresponding to the minimum of phase
sensitivity, we plot the phase sensitivity $\Delta \phi $ as a function of $%
\phi $ for several different values $\left \vert \gamma \right \vert
=0,1,2,3 $, as shown in Fig. 3(a). Significantly, for all given values $%
\left \vert \gamma \right \vert =0,1,2,3$, the minimum of phase sensitivity
can be always found at the optimal point $\phi =0$. In addition, it is more
interesting that, when fixed the optimal point $\phi =0$, as the LDO
strength $\left \vert \gamma \right \vert $ increases, the corresponding
value of phase sensitivity can further decrease, which fully demonstrates
that the usage of the LDO in SU(1,1) interferometer systems is more
beneficial for improving the phase sensitivity than that without the LDO.
The reason for these phenomena is that the increase of the LDO strength $%
\left \vert \gamma \right \vert $ contributes to increase the slope $%
\partial \left \langle \hat{X}\right \rangle /\partial \phi $\ of the output
signal $\left \langle \hat{X}\right \rangle $, thereby giving rise to the
increase of the denominator in Eq. (\ref{8}), as pictured in Fig 3(b).
\begin{figure*}[tbp]
\label{Fig4} \centering \includegraphics[width=0.72\columnwidth]{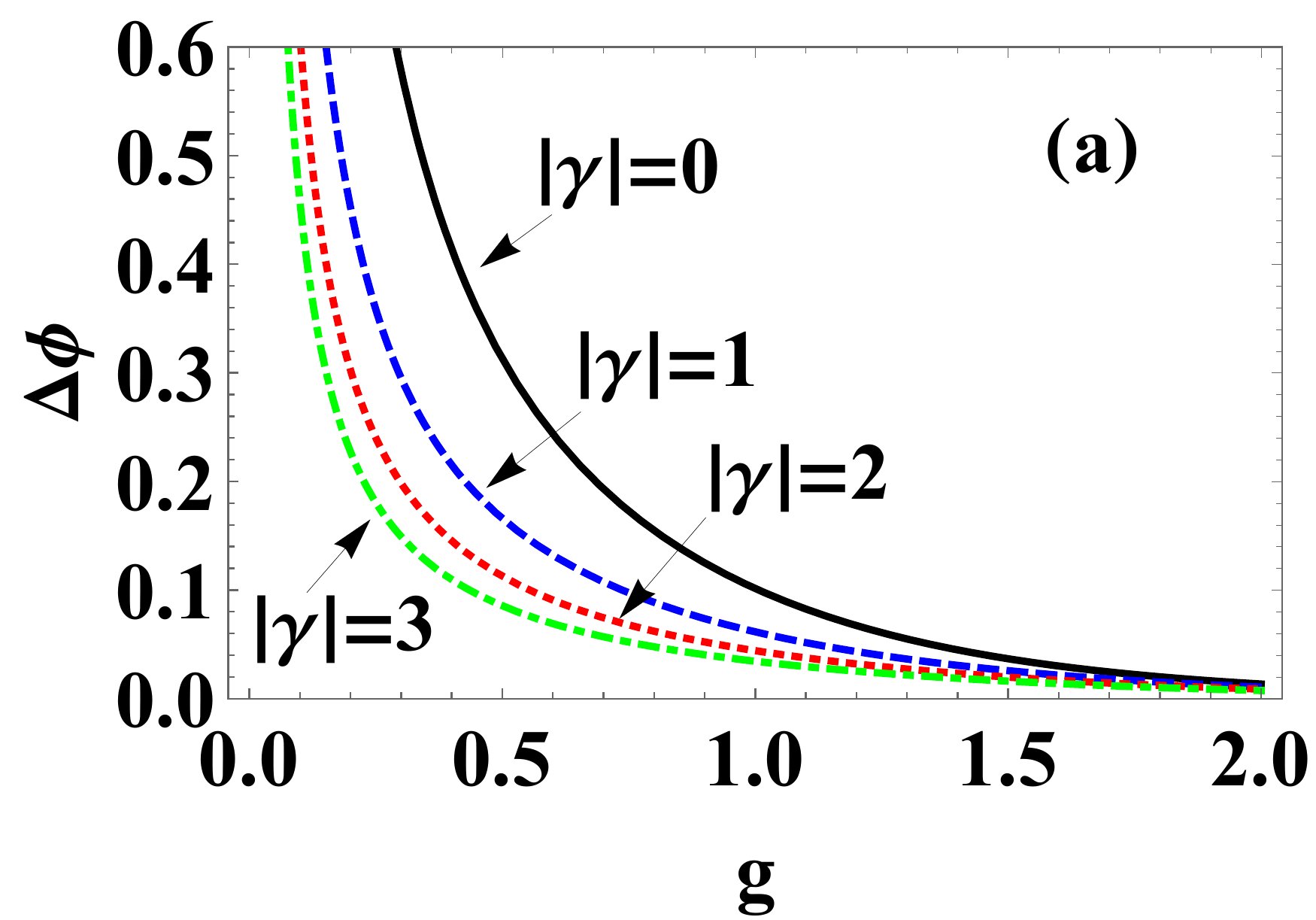}%
\includegraphics[width=0.72\columnwidth]{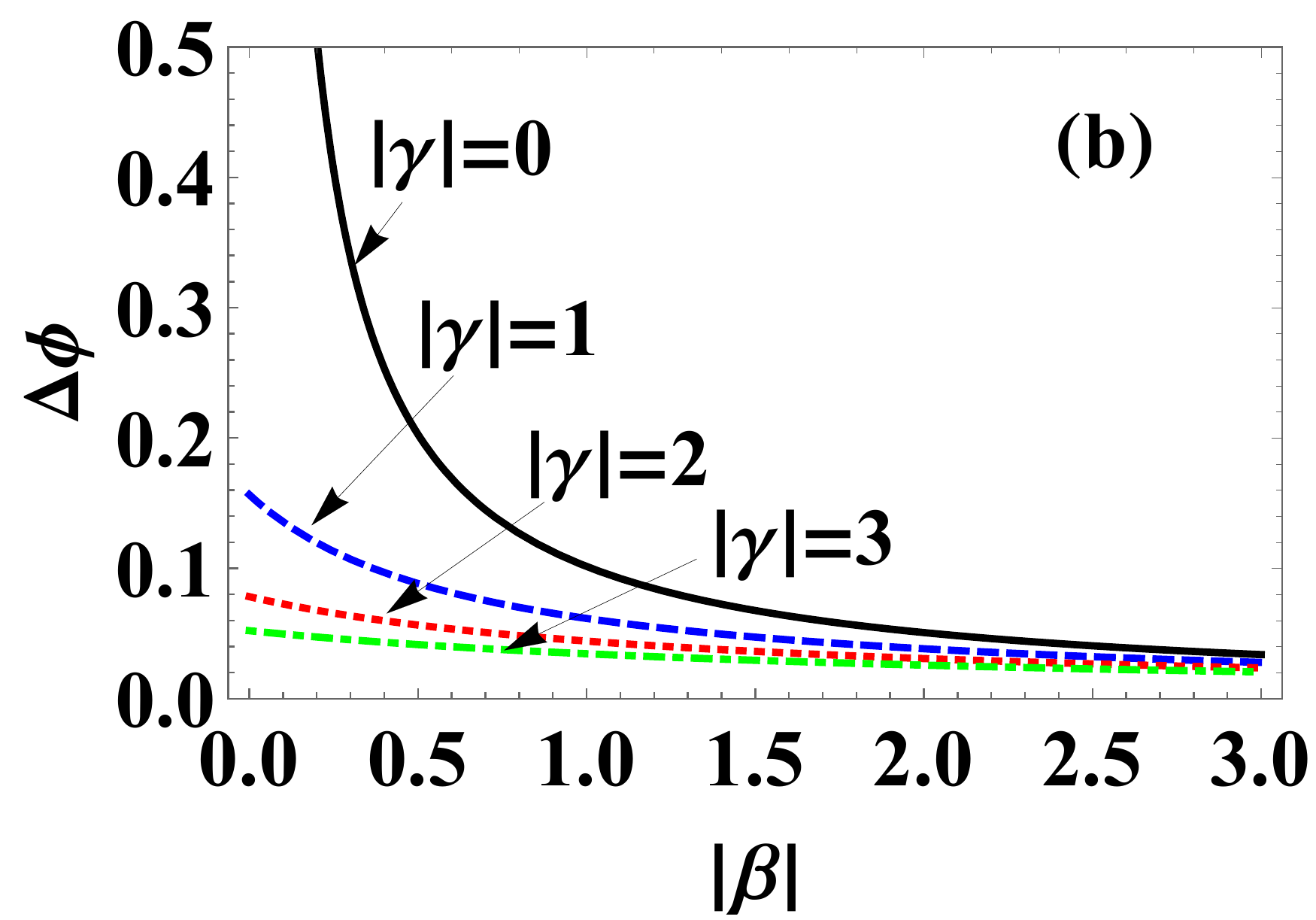} %
\includegraphics[width=0.72\columnwidth]{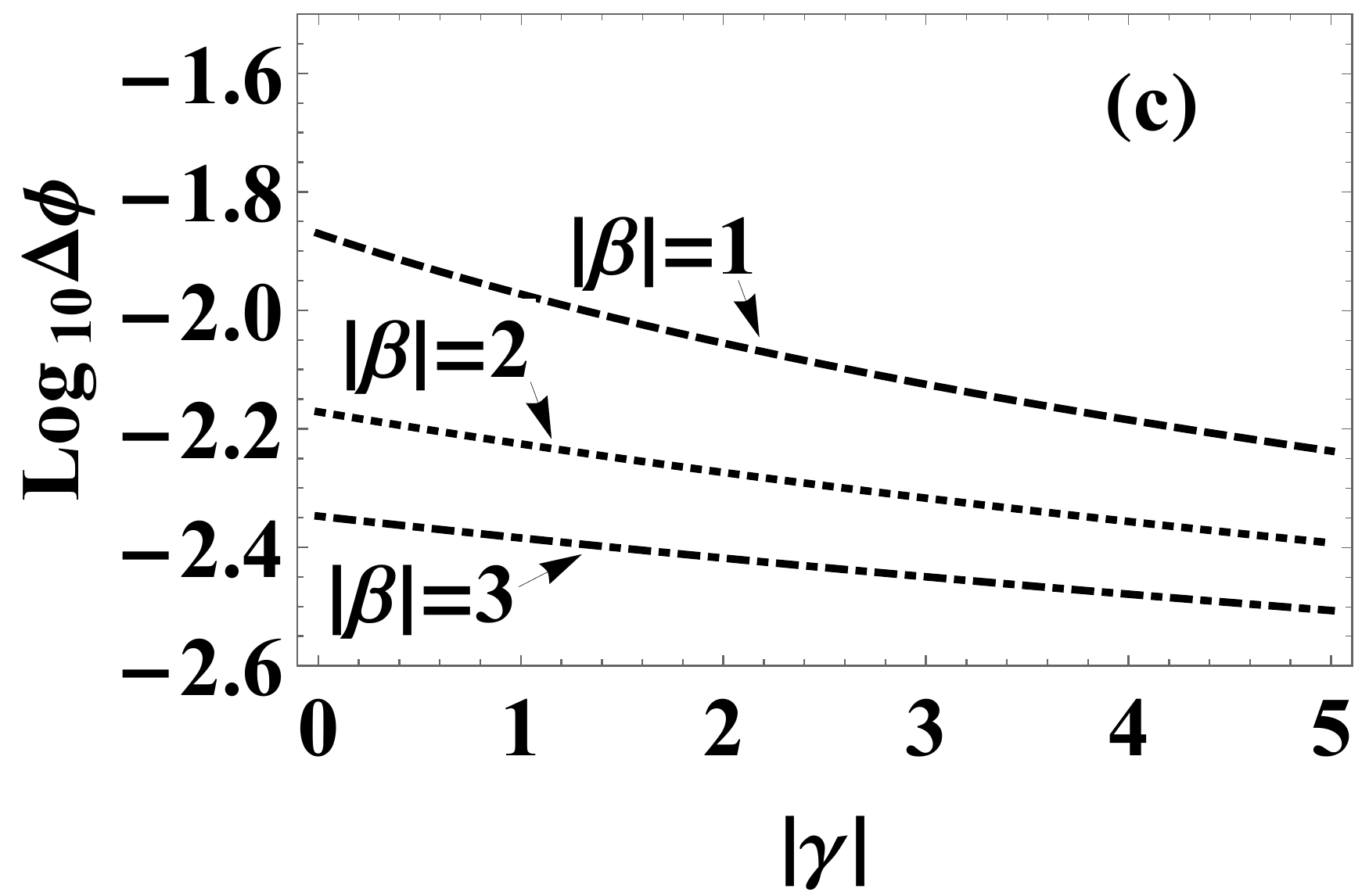} %
\includegraphics[width=0.72\columnwidth]{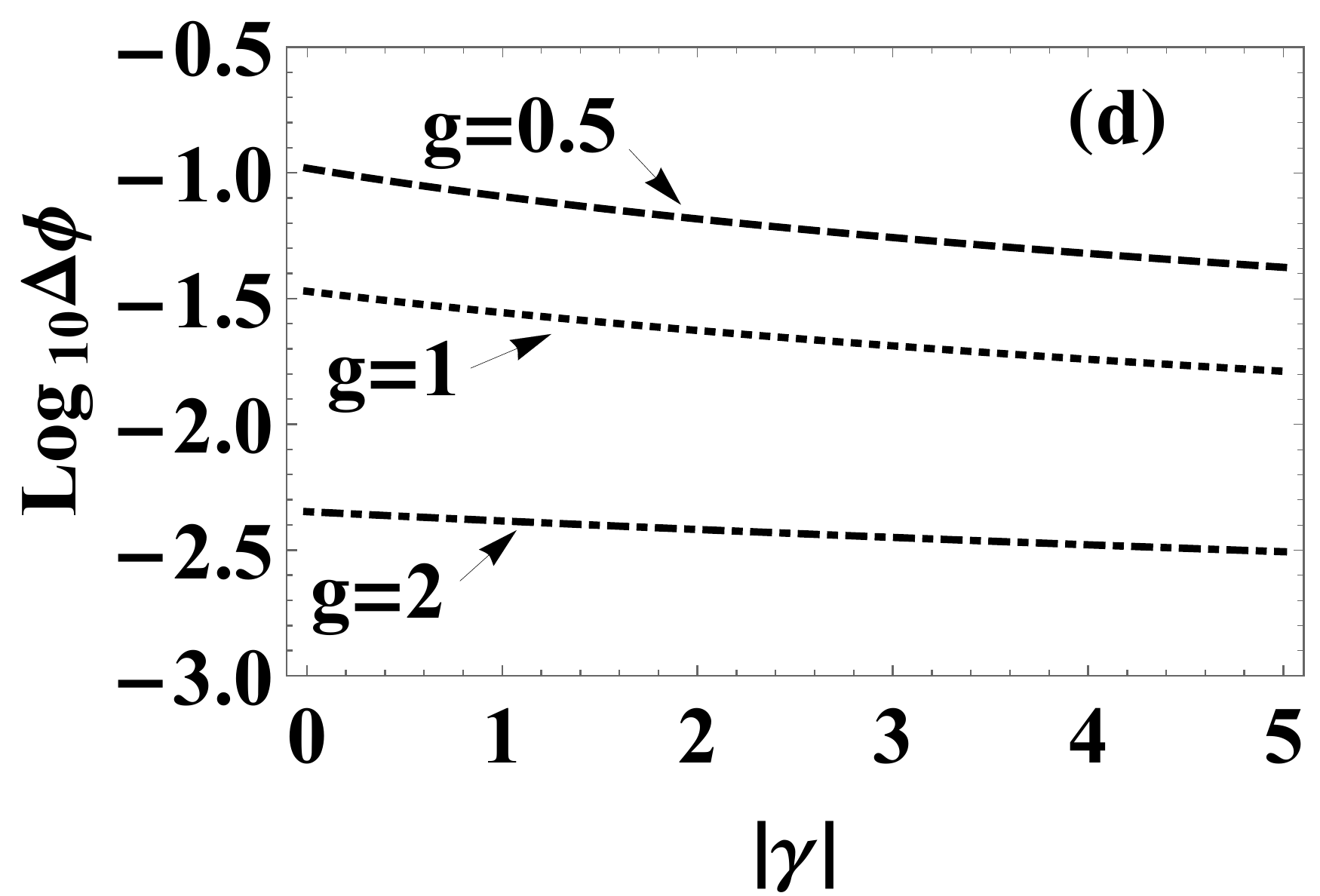}
\caption{(Color online) Phase sensitivity changing with (a) $g$, (b) $%
\left
\vert \protect \beta \right \vert $ and (c)-(d) $\left \vert \protect%
\gamma \right \vert $. Among them, in(a), the fixed values are $\left \vert
\protect \beta \right \vert =r=1$ for different $\left \vert \protect \gamma %
\right
\vert =0,1,2,3$; in (b), the fixed values are $g=r=1$ for different $%
\left
\vert \protect \gamma \right \vert =0,1,2,3$; in (c), the fixed values
are $g=2,r=1$ for different $\left \vert \protect \beta \right \vert =1,2,3$;
in (d), the fixed values are $\left \vert \protect \beta \right \vert =3$ $%
,r=1$ for different $g=0.5,1,2$. Other parameters are as following: $\protect%
\phi =\protect \theta _{\protect \xi }=0$ and $\protect \theta _{\protect \beta %
}=\protect \theta _{\protect \gamma }=\protect \pi /2$.}
\end{figure*}
After achieving the optimal point $\phi =0$, we can further obtain the phase
sensitivity $\Delta \phi $ of DSU(1,1) interferometer around this optimal
point, i.e.,%
\begin{equation}
\left. \Delta \phi \right \vert _{\phi =0}=\frac{e^{-r}}{2\left \vert \beta
\right \vert \sinh g\cosh g+2\left \vert \gamma \right \vert \sinh g},
\label{9}
\end{equation}%
where the second term of the denominator results from the LDO. In
particular, when $\left \vert \gamma \right \vert =0$ corresponding to the
SU(1,1) interferometer without the LDO, its phase sensitivity $\Delta \phi $
is consistent with the previous work \cite{17}. To visually compare the
phase sensitivity of SU(1,1) interferometers with and without the LDO, Fig.
4 shows the phase sensitivity changing with $g$, $\left \vert \beta \right
\vert $ and $\left \vert \gamma \right \vert $. In Figs. 4(a) and 4(b), the
black solid line indicates the phase sensitivity of SU(1,1) interferometer
without the LDO, which can be exceeded by that of DSU(1,1) interferometer ($%
\left \vert \gamma \right \vert \neq 0$), and the phase sensitivity of
SU(1,1) interferometer systems can be further enhanced via the increase of
the LDO strength $\left \vert \gamma \right \vert $. The reason may be that
the QFI can be effectively \ increased by increasing the values of $\left
\vert \gamma \right \vert $, which can be seen in Figs. 2(a) and 2(b).
Moreover, we also find that, for given a LDO strength $\left \vert \gamma
\right \vert $, the corresponding phase sensitivity can be further improved
by increasing the values of $g$ and $\left \vert \beta \right \vert $, but
for sufficiently large $g$ and $\left \vert \beta \right \vert $, the
improvement effects can not be clearly distinguished. For this reason, when
given $g=2$ and $\left \vert \beta \right \vert =3$, we also respectively
plot the $\log _{10}\Delta \phi $ versus $\left \vert \gamma \right \vert $,
as shown in Figs. 4(c) and 4(d). Obviously, with the increase of $\left
\vert \gamma \right \vert $, the improved effects of phase sensitivity are
still evident. For instance, at a fixed $g=2$, the corresponding phase
sensitivity for $\left \vert \beta \right \vert =1$ can be further improved
by about $57\%$ when $\left \vert \gamma \right \vert $ ranges from $0$ to $5
$, see the black dashed line in Fig. 4(c).
\begin{figure}[tbp]
\label{Fig5} \centering \includegraphics[width=0.72\columnwidth]{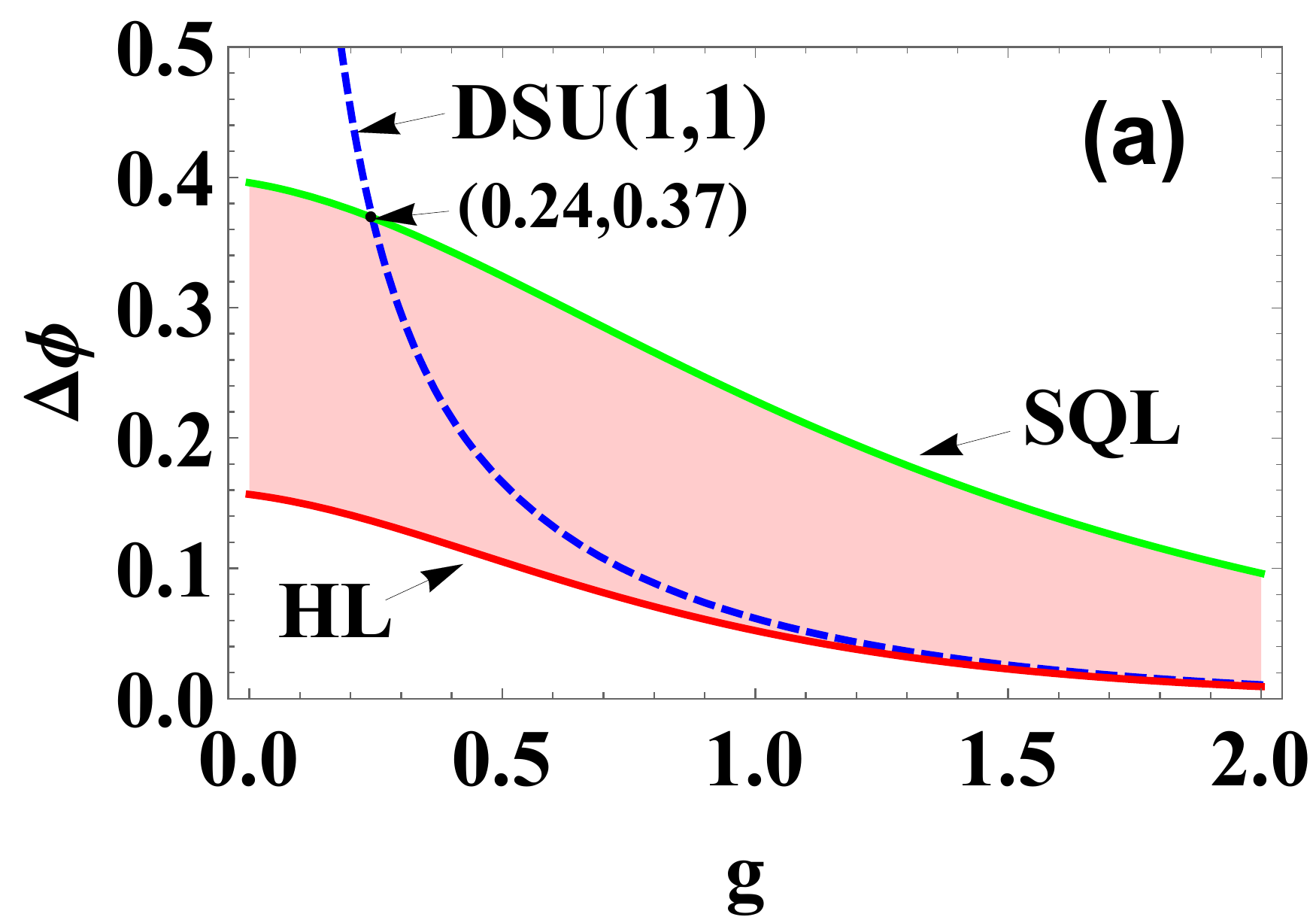}
\newline
\includegraphics[width=0.73\columnwidth]{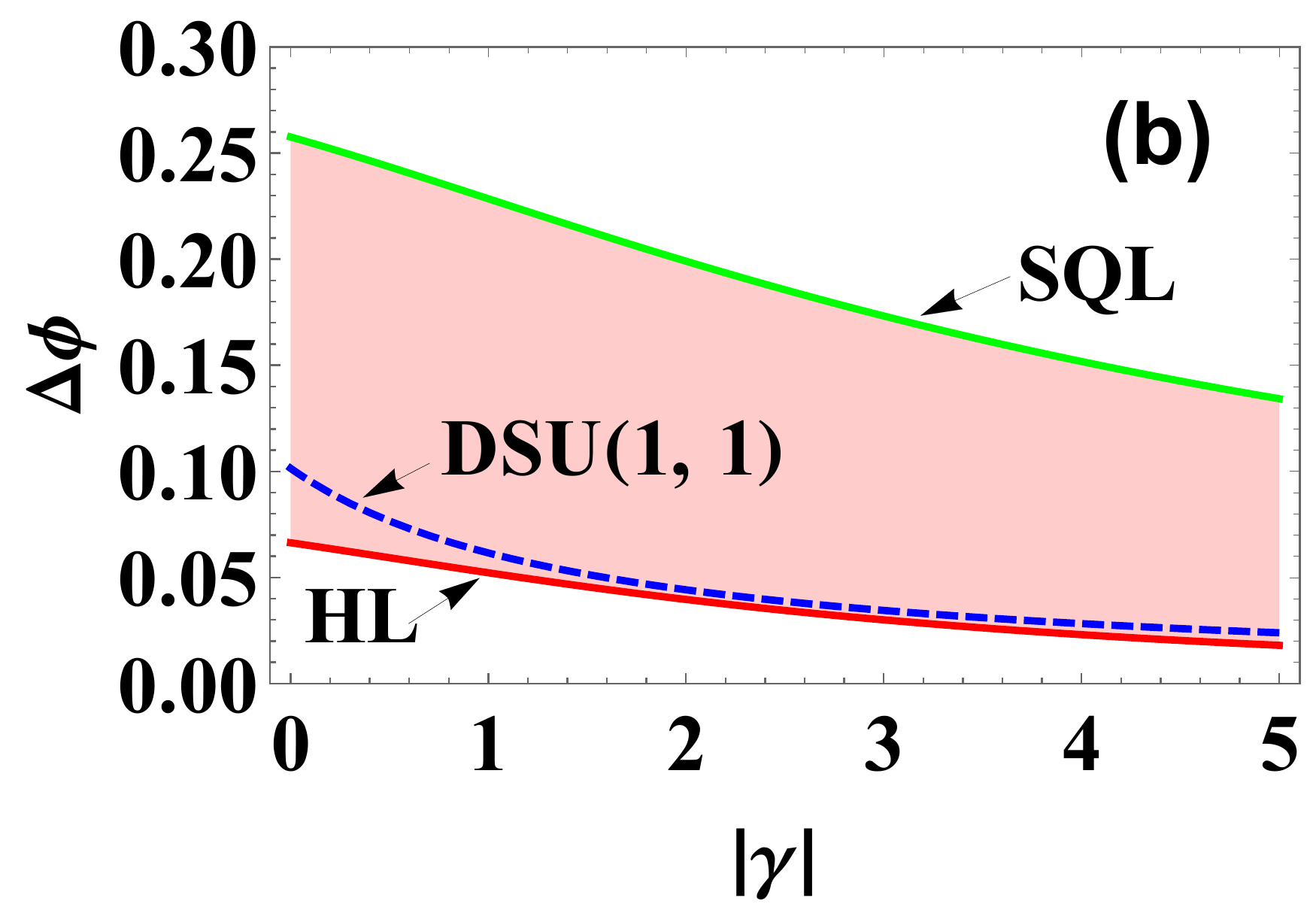} \newline
\caption{(Color online) Comparison about precision limits involving the SQL,
the HL and the SU(1,1) interferometer with the LDOs. (a) and (b)
respectively corresponds to phase sensitivity $\Delta \protect \phi $
changing with $g$ and $|\protect \gamma |$, when fixed $|\protect \gamma |=|%
\protect \beta |=r=1$ and $|\protect \beta |=g=r=1$. }
\end{figure}

\section{Phase sensitivity of DSU(1,1) interferometer with the SQL and the HL%
}

After presenting both the QFI and the phase sensitivity of DSU(1,1)
interferometer in an ideal scenario, we shall make a comparison about phase
sensitivities including the SQL, the HL and the DSU(1,1) interferometer
scheme. For this purpose, the total mean photon number inside DSU(1,1)
interferometer should be introduced, which can be defined as
\begin{equation}
N_{Total}=\left \langle \psi _{\gamma }\right \vert (\hat{a}^{\dagger }\hat{a%
}+\hat{b}^{\dagger }\hat{b})\left \vert \psi _{\gamma }\right \rangle ,
\label{10}
\end{equation}%
where $\left \vert \psi _{\gamma }\right \rangle =\hat{D}_{a}(\gamma )\hat{D}%
_{b}(\gamma )\hat{U}_{OPA1}\left \vert \psi _{in}\right \rangle $ is the
probe state just after the LDO with both $\hat{U}_{OPA1}=\exp
(g_{1}e^{-i\theta _{1}}\hat{a}\hat{b}-g_{1}e^{i\theta _{1}}\hat{a}^{\dagger }%
\hat{b}^{\dagger })$ being the OPA$_{1}$ process and $\left \vert \psi
_{in}\right \rangle =\left \vert \xi \right \rangle _{a}\otimes \left \vert
\beta \right \rangle _{b} $ being the input state of DSU(1,1)
interferometer. It is worth noting that the total mean photon number $%
N_{Total}$ inside DSU(1,1) interferometer is different from the total mean
photon number $\bar{N}_{in}=\left \vert \beta \right \vert ^{2}+\sinh ^{2}r$
of the input state $\left \vert \psi _{in}\right \rangle $, which in our
scheme can be given by%
\begin{eqnarray}
&&N_{Total}=\bar{N}_{in}\cosh 2g+2\sinh ^{2}g  \notag \\
&&+2\left \vert \gamma \right \vert \left \vert \beta \right \vert (\cosh
g+\sinh g)+2\left \vert \gamma \right \vert ^{2},  \label{11}
\end{eqnarray}%
where the first two terms result from the amplification process of $\bar{N}%
_{in}$ and the spontaneous process prior to the implementation of the LDO,
and the last two terms stem from the LDO process. According to Eq. (\ref{11}%
), one can respectively derive the SQL and the HL, i.e.,

\begin{eqnarray}
\Delta \phi _{SQL} &=&\frac{1}{\sqrt{N_{Total}}},  \notag \\
\Delta \phi _{HL} &=&\frac{1}{N_{Total}}.  \label{12}
\end{eqnarray}%
\  \  \  \  \  \ To see if the phase precision of DSU(1,1) interferometer can
surpass the SQL, even closing to the HL, we make a comparison about the
phase sensitivity changing with $g$ and $\left \vert \gamma \right \vert $,
as shown in Fig. 5(a) and 5(b). It is clearly seen from Fig. 5(a) that, at
fixed values of $\left \vert \beta \right \vert =$ $\left \vert \gamma
\right \vert =r=1$, for a large range of $g$ (i.e., $g>0.24$), the phase
precision of DSU(1,1) interferometer scheme can easily break through the SQL
(solid green line), even gradually approaching to HL (solid red line) with
the increase of $g$. In addition, when given $g=\left \vert \beta
\right
\vert =r=1$, it is found in Fig. 5(b) that the phase precision of
DSU(1,1) interferometer scheme is alway superior to the SQL, and can
gradually approach to HL when increasing the value of $\left \vert \gamma
\right \vert $, which implies that the applications of the LDO into the
SU(1,1) interferometer systems are salutary aspects to the enhancement of
phase precision, making it close to the HL.

\section{The QFI and phase sensitivity of the photon-loss DSU(1,1)
interferometer}

For the realistic environment, the photon losses are of vital importance in
restricting the precision of quantum metrology. In particular, how the
photon losses affect the quantum-noise cancellation in SU(1,1)
interferometer was discussed in detail both theoretically and experimentally
\cite{53}. As a result, in this section, we shall analyze and discuss the
behaviors of both the QFI and the phase sensitivity in the photon-loss
DSU(1,1) interferometer.
\begin{figure*}[tbp]
\label{Fig6} \centering
\subfigure[]{
\centering
\includegraphics[width=0.72\columnwidth]{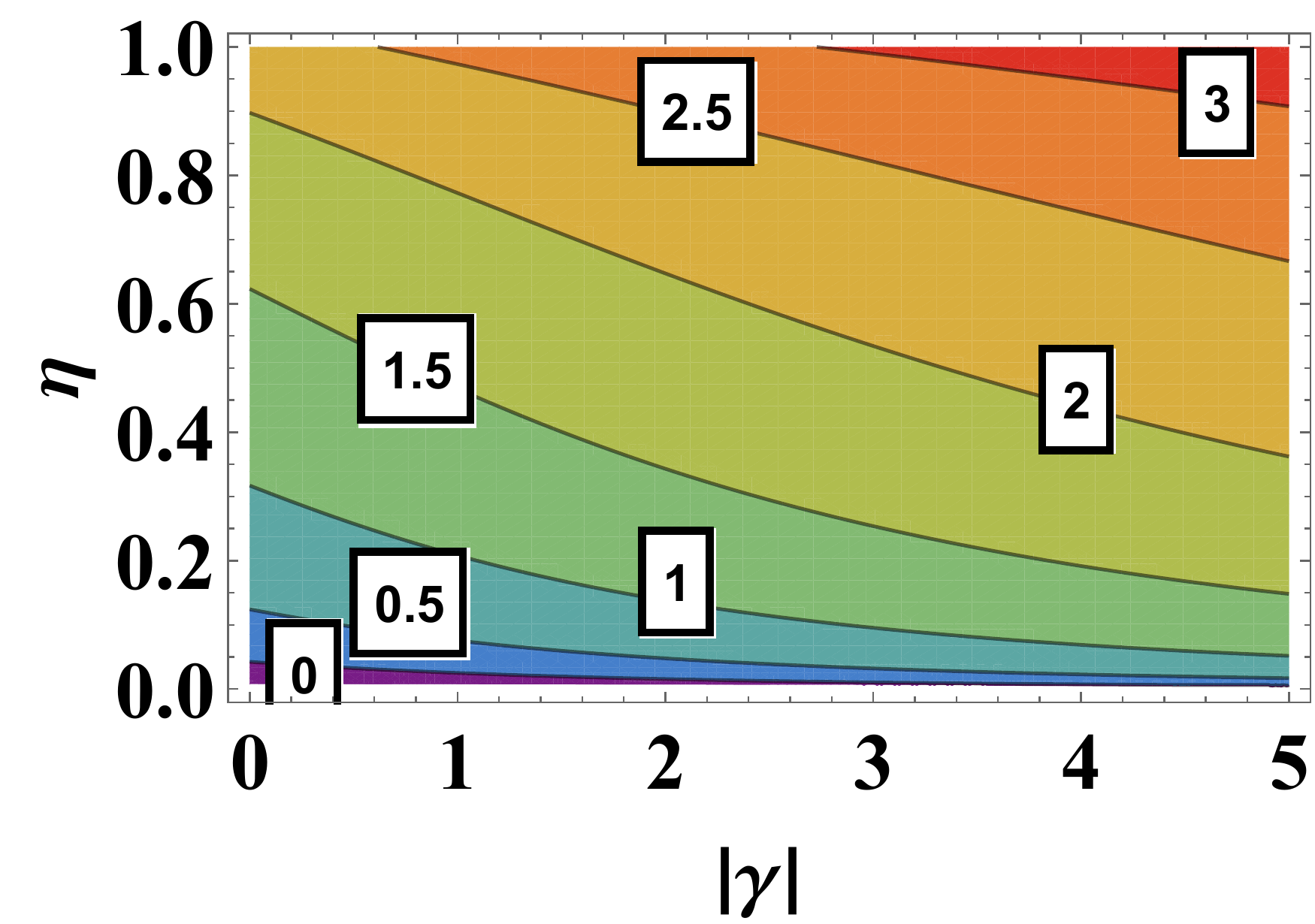}
\label{Fig6a}
}
\subfigure[]{
\includegraphics[width=0.72\columnwidth]{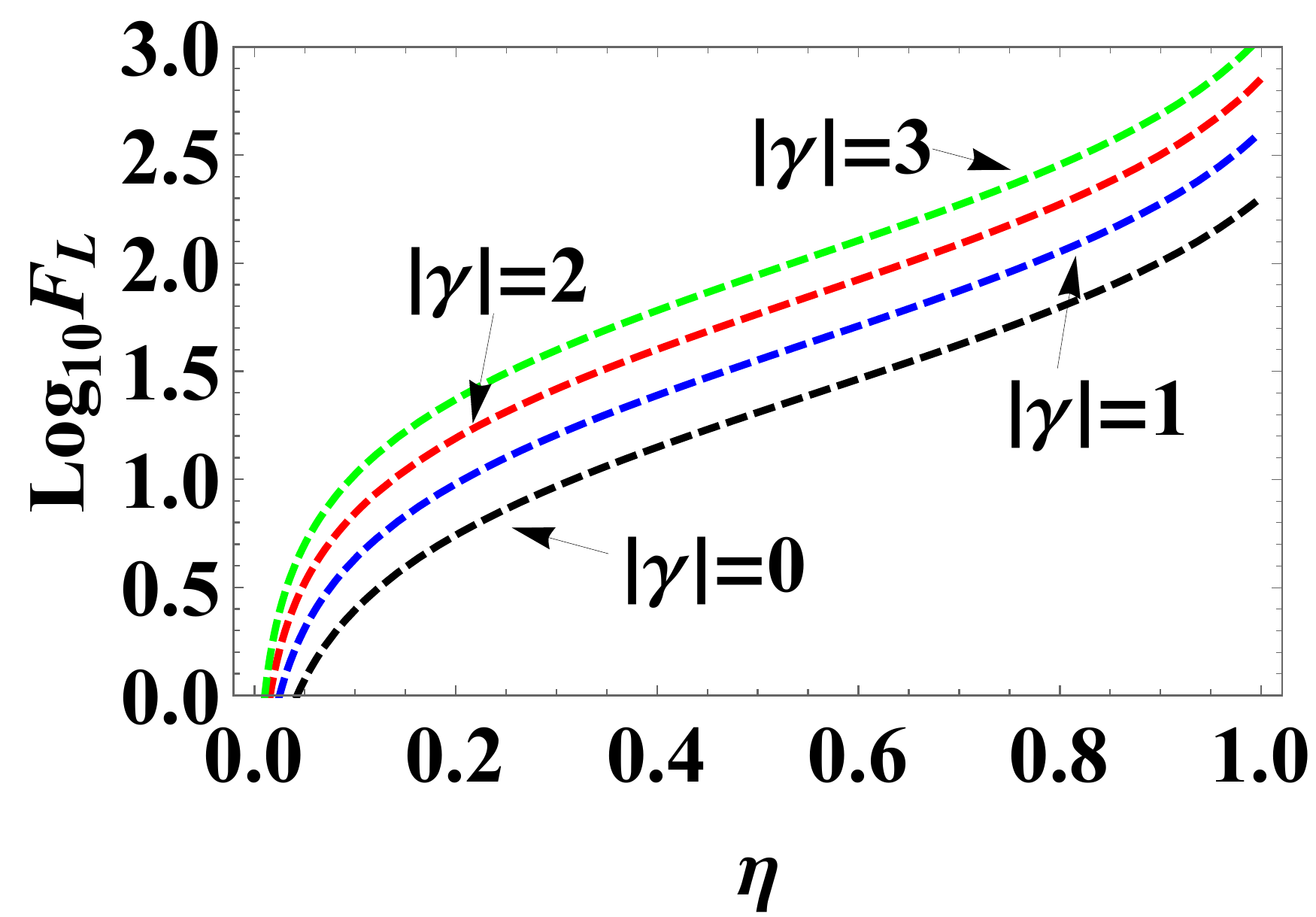}
\label{Fig6b}
}
\subfigure[]{
\includegraphics[width=0.72\columnwidth]{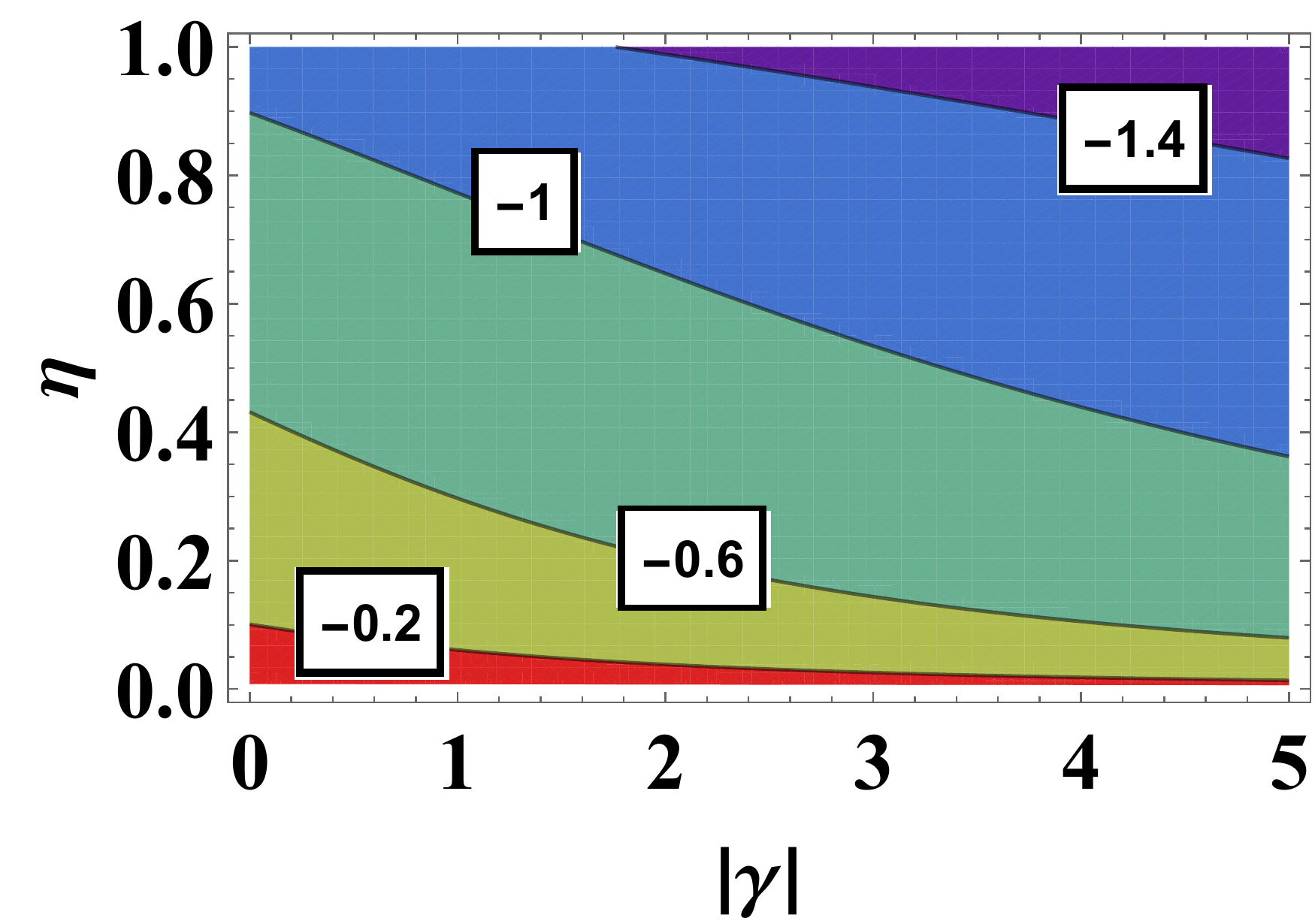}
\label{Fig6c}
}
\subfigure[]{
\includegraphics[width=0.72\columnwidth]{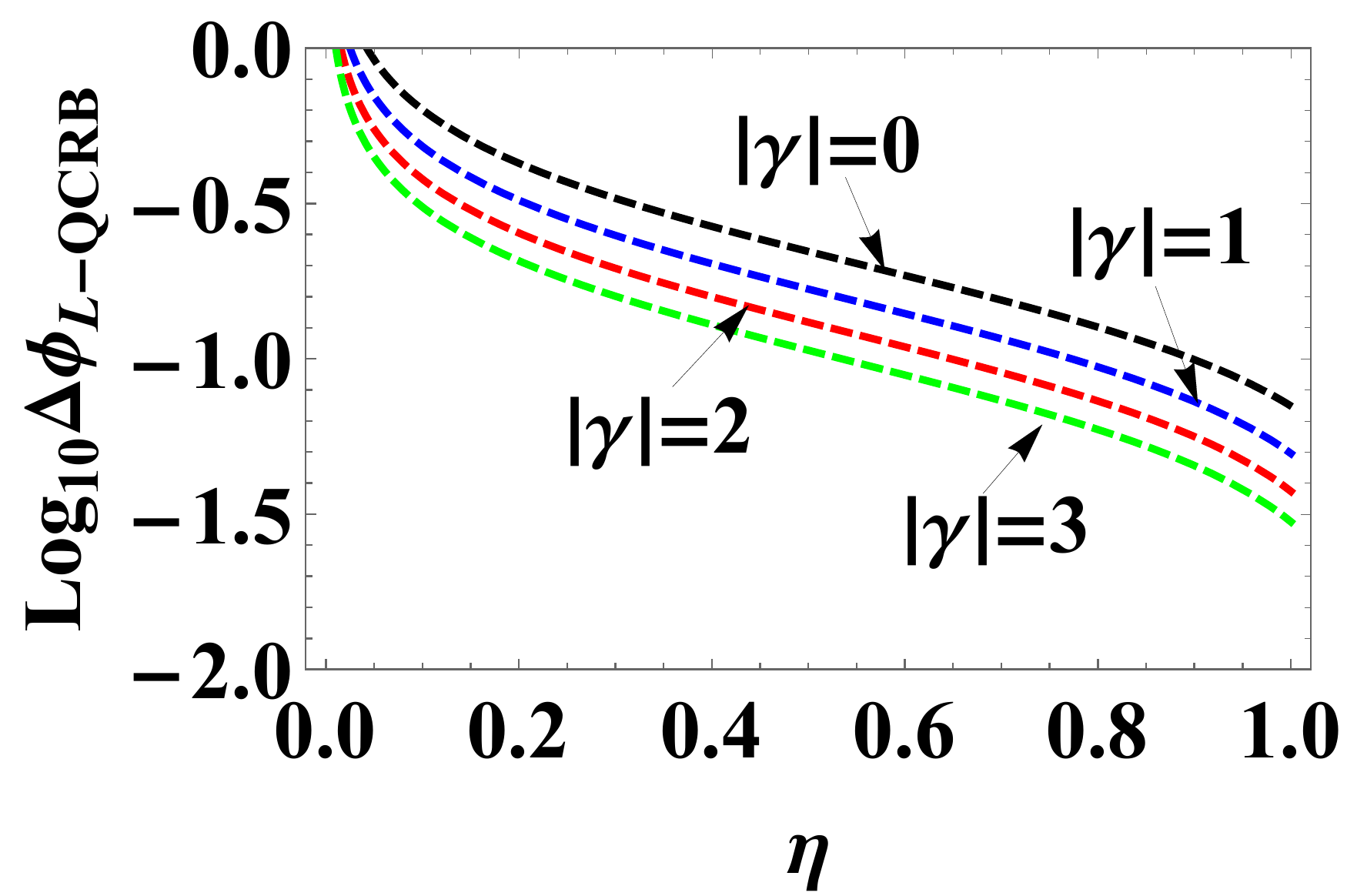}
\label{Fig6d}
}
\caption{ (Color online) The performances of both (a)-(b) the QFI and
(c)-(d) the QCRB under the photon losses. Among them, the counterplots of
both (a) the QFI $\log _{10}F$ and (c) the QCRB $\log _{10}\Delta \protect%
\phi _{QCRB}$ in ($\protect \eta ,\left \vert \protect \gamma \right \vert $)
space, whereas (b) the QFI $\log _{10}F$ and (d) the QCRB $\log _{10}\Delta
\protect \phi _{QCRB}$ as a function of $\protect \eta $ for several different
$\left \vert \protect \gamma \right \vert =0,1,2,3$. Other parameters are as
following: $\protect \phi $=$\protect \theta _{\protect \xi }$=$0$, $%
\left
\vert \protect \beta \right \vert $=$g$=$r$=$1$ and $\protect \theta _{%
\protect \beta } $=$\protect \theta _{\protect \gamma }$=$\protect \pi /2$. }
\end{figure*}

\subsection{The effects of photon losses on the QFI}

Due to the existence of photon losses, it is not suitable for deriving the
QFI via the conventional method given in Eq. (\ref{4}). To solve this
problem, a novel variational method was proposed by Escher \cite{54}, which
has been used in the photon-loss single-(or multi-)parameter estimation
systems \cite{22,55,56}. Inspired by this lightspot, below we would derive
the explicit form of the QFI in the photon-loss DSU(1,1) interferometer,
with the help of the variational method.

Originally, we first denote the probe state as $\left \vert \psi _{\gamma
}\right \rangle \equiv \left \vert \psi _{\gamma }\right \rangle _{S}$ where
$\left \vert \psi _{\gamma }\right \rangle _{S}$ is an initial probe state
of DSU(1,1) interferometer system $S$. Because of the photon losses, the
encoding process of the probe state $\left \vert \psi _{\gamma
}\right
\rangle $ to an unknown phase $\phi $ is no longer the unitary
evolution, so that the system $S$ is expanded into the enlarged one $S+E$ ($%
E $ represents the photon-loss environment system). Under such
circumstances, the initial probe state $\left \vert \psi _{\gamma
}\right
\rangle _{S}$ in the enlarged systems $S+E$ experiences the unitary
phase encoding process $\hat{U}_{S+E}(\phi )$, which can be described as
\cite{54}
\begin{eqnarray}
\left \vert \psi \right \rangle _{S+E} &=&\hat{U}_{S+E}(\phi )\left \vert
\psi _{\gamma }\right \rangle _{S}\left \vert 0\right \rangle _{E}  \notag \\
&=&\sum_{j=0}^{\infty }\hat{K}_{j}(\phi )\left \vert \psi _{\gamma }\right
\rangle _{S}\left \vert j\right \rangle _{E},  \label{13}
\end{eqnarray}%
where $\left \vert 0\right \rangle _{E}$ is the initial state of the
photon-loss system $E,$ $\left \vert j\right \rangle _{E}$ is the orthogonal
basis of the $\left \vert 0\right \rangle _{E},$ and $\hat{K}_{j}(\phi )$ is
the Kraus operator only working on the $\left \vert \psi _{\gamma
}\right
\rangle _{S}$, whose expression can be given by \cite{54}
\begin{equation}
\hat{K}_{j}(\phi )=\sqrt{\frac{(1-\eta )^{j}}{j!}}e^{i\phi (\hat{b}^{\dagger
}\hat{b}-\lambda j)}\eta ^{\hat{b}^{\dagger }\hat{b}/2}\hat{b}^{j},
\label{14}
\end{equation}%
with the variational parameter $\lambda $ and the strength of the photon
losses $\eta $ ($\eta =0$ and $\eta =1$ respectively denote the complete
absorption and lossless cases). In this situation, the QFI for the DSU(1,1)
interferometer with the photon losses can be given by \cite{54,55,57}
\begin{figure*}[tbp]
\label{Fig7} \centering
\subfigure[]{
\centering
\includegraphics[width=0.72\columnwidth]{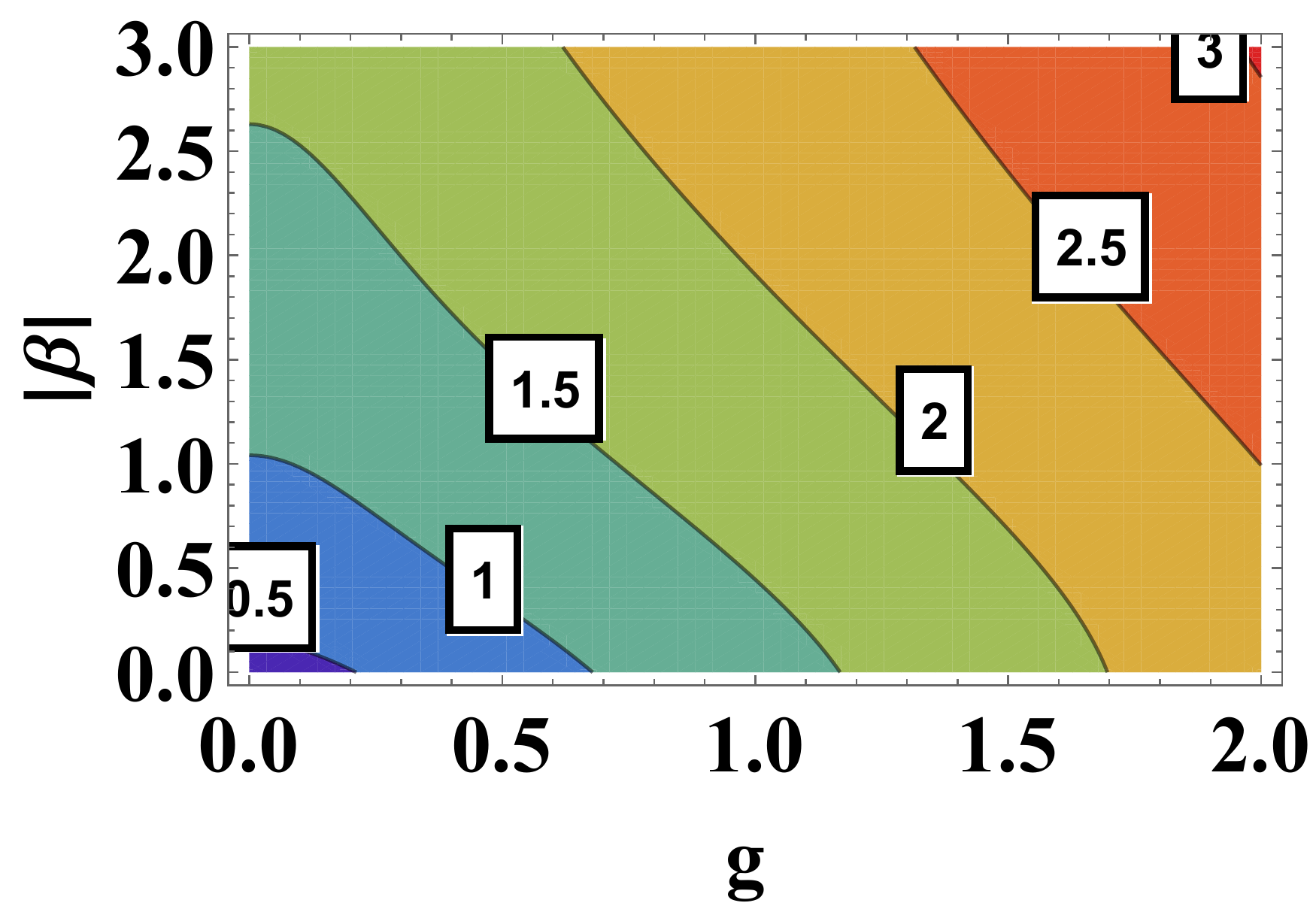}
\label{Fig7a}
}
\subfigure[]{
\includegraphics[width=0.72\columnwidth]{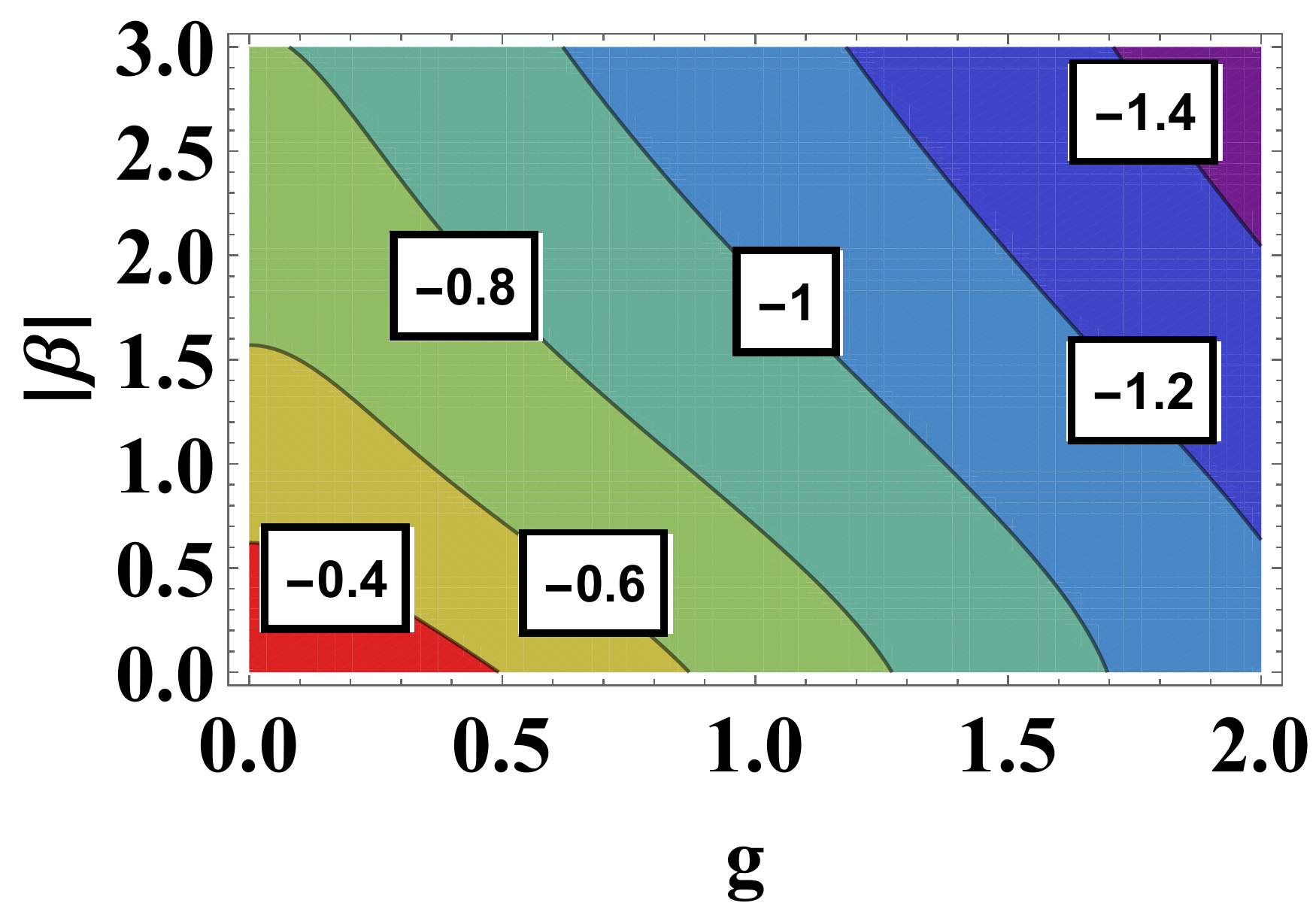}
\label{Fig7b}
}
\caption{ (Color online) The counterplots of both (a) the QFI $\log _{10}F$
and (b) the QCRB $\log _{10}\Delta \protect \phi _{QCRB}$ in ($\left \vert
\protect \beta \right \vert ,g$) space. Other parameters are as following: $%
\protect \phi $=$\protect \theta _{\protect \xi }$=$0$, $\left \vert \protect%
\eta \right \vert $=$0.6$, $\left \vert \protect \gamma \right \vert $=$r$=$1$
and $\protect \theta _{\protect \beta }$=$\protect \theta _{\protect \gamma }$=$%
\protect \pi /2$. }
\end{figure*}
\begin{figure}[tbp]
\label{Fig8} \centering \includegraphics[width=0.72\columnwidth]{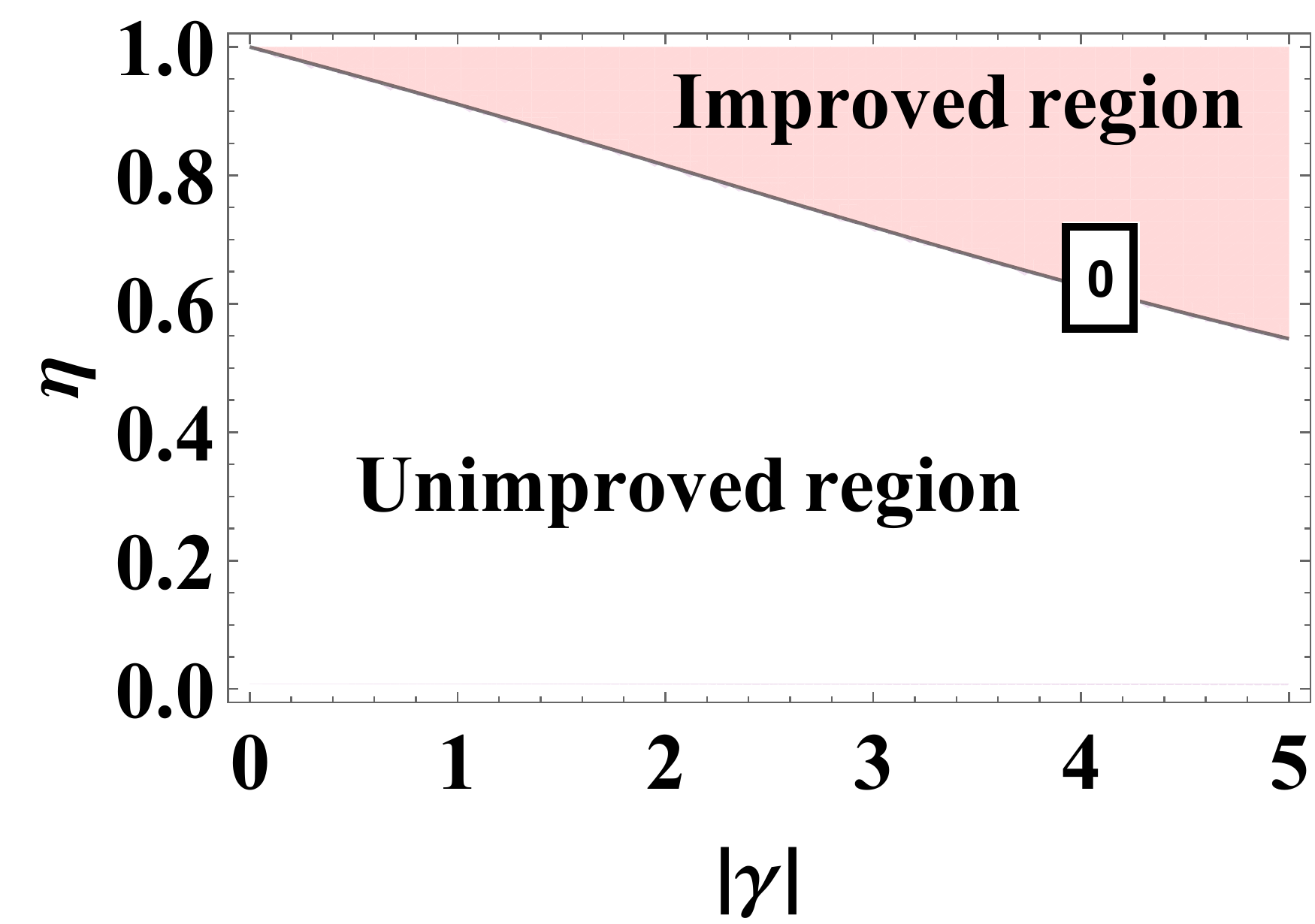}
\caption{{}(Color online) The difference between the QCRB with the photon
losses and the one without both the LDOs and the photon losses the QCRB $%
\widetilde{\Delta }$=$\log _{10}\Delta \protect \phi _{L-QCRB}-\log
_{10}\Delta \protect \phi _{QCRB}$ in ($\left \vert \protect \eta \right \vert
,\protect \gamma $) space. Note that the color region represents the
condition of $\widetilde{\Delta }<0$. Other parameters are as following: $%
\protect \phi $=$\protect \theta _{\protect \xi }$=$0$, $|\protect \beta |$=$g$=$%
r$=$1$ and $\protect \theta _{\protect \beta }$=$\protect \theta _{\protect%
\gamma }$=$\protect \pi /2$. }
\end{figure}
\begin{equation}
F_{L}=\min_{\left \{ \hat{K}_{j}(\phi )\right \} }C_{Q}[\left \vert \psi
_{\gamma }\right \rangle _{S},\hat{K}_{j}(\phi )],  \label{15}
\end{equation}%
with the upper bound of the QFI in the photon-losses systems \cite{54}
\begin{eqnarray}
C_{Q}[\left \vert \psi _{\gamma }\right \rangle _{S},\hat{K}_{j}(\phi )] &=&4%
\left[ _{S+E}\left \langle \psi ^{\prime }|\psi ^{\prime }\right \rangle
_{S+E}\right.  \notag \\
&&\left. -|_{S+E}\left \langle \psi ^{\prime }|\psi \right \rangle
_{S+E}|^{2} \right] .  \label{16}
\end{eqnarray}%
Upon substituting Eqs. (\ref{13}) into (\ref{16}), so that
\begin{equation}
C_{Q}[\left \vert \psi _{\gamma }\right \rangle _{S},\hat{K}_{j}(\phi )]=4%
\left[ \left \langle \hat{H}_{1}(\phi )\right \rangle _{S}-\left \langle
\hat{H}_{1}(\phi )\right \rangle _{S}^{2}\right] ,  \label{17}
\end{equation}%
where the symbol of $\left \langle \cdot \right \rangle $ is the inner
product with respect to the initial probe state $\left \vert \psi _{\gamma
}\right \rangle _{S}$ and%
\begin{eqnarray}
\hat{H}_{1}(\phi ) &=&\sum_{j=0}^{\infty }\frac{d\hat{K}_{j}^{\dagger }(\phi
)}{d\phi }\frac{d\hat{K}_{j}(\phi )}{d\phi },  \notag \\
\hat{H}_{2}(\phi ) &=&i\sum_{j=0}^{\infty }\frac{d\hat{K}_{j}^{\dagger
}(\phi )}{d\phi }\hat{K}_{j}(\phi ).  \label{18}
\end{eqnarray}%
Thus, combining Eqs. (\ref{14}) and (\ref{17}), Eq. (\ref{16}) can be
rewritten as \cite{54}
\begin{eqnarray}
C_{Q}[\left \vert \psi _{\gamma }\right \rangle _{S},\hat{K}_{j}(\phi )]
&=&4(\eta +\eta \lambda -\lambda )^{2}\left \langle \Delta ^{2}\hat{n}\right
\rangle  \notag \\
&&+4\eta (1-\eta )(1+\lambda )^{2}\left \langle \hat{n}\right \rangle ,
\label{19}
\end{eqnarray}%
with the symbol of $\left \langle \Delta ^{2}\cdot \right \rangle $
representing the variance of the $\left \vert \psi _{\gamma }\right \rangle
_{S}$. From Eq. (\ref{19}), when minimizing the upper bound of the QFI, the
optimal value of $\lambda $ is calculated as \cite{54}
\begin{equation}
\lambda _{opt}=\frac{\left \langle \Delta ^{2}\hat{n}\right \rangle }{%
(1-\eta )\left \langle \Delta ^{2}\hat{n}\right \rangle +\eta \left \langle
\hat{n}\right \rangle }-1,  \label{20}
\end{equation}%
so that according to Eqs. (\ref{15}) and (\ref{20}), the explicit form of
the QFI with the photon losses can be finally derived as \cite{22,54}%
\begin{equation}
F_{L}=\frac{4\eta F\left \langle \hat{n}\right \rangle }{(1-\eta )F+4\eta
\left \langle \hat{n}\right \rangle },  \label{21}
\end{equation}%
where $F$ corresponds to the lossless case given in Eq. (\ref{5}). For our
scheme, when considering the $\left \vert \xi \right \rangle _{a}\otimes $ $%
\left \vert \beta \right \rangle _{b}$ as the inputs of DSU(1,1)
interferometer, Eq. (\ref{21}) can be rewritten as
\begin{equation}
F_{L}=\frac{4\eta F\Gamma _{1}}{(1-\eta )F+4\eta \Gamma _{1}}.  \label{22}
\end{equation}%
It is clearly seen from Eq. (\ref{22}) that when $\eta =1$, one can obtain $%
F_{L}=$ $F$ corresponding to the QFI with the ideal case. To intuitively
elaborate the effects of the photon losses on the estimation performance of
DSU(1,1) interferometer, we show the behaviors of both the QFI $\log
_{10}F_{L}$\ and the QCRB $\log _{10}$\ $\Delta \phi _{L-QCRB}$\ changing
with $\eta \ $and $\left \vert \gamma \right \vert $, as depicted in Fig. 6.
As can be seen from Figs. 6(a), when given a certain value of $\left \vert
\gamma \right \vert $, the value of the QFI decreases with the decrease of $%
\eta $, which means that the QFI is heavily influenced by the strength of
the photon losses $\eta $. Even so, we also can find that, at a fixed $\eta $%
, e.g., $\eta =0.6$, the QFI can be further improved by increasing the value
of $\left \vert \gamma \right \vert $, implying that the usage of the LDO is
conducive to resisting the photon losses, thereby achieving the higher phase
sensitivity of SU(1,1) interferometer systems, which can be seen in Fig.
6(b). These results are also true for the QCRB, as shown in Figs. 6(c) and
6(d). More interestingly, at fixed $\eta =0.6$\ and $\left \vert \gamma
\right \vert =1$\ , it is also possible to further enhance both the QFI and
the QCRB via the increasing parameters of $g$\ and $\left \vert \beta
\right
\vert $, as seen in Fig. 7.

Finally, to show the advantages of exploiting the LDO into the SU(1,1)
interferometer, we take the difference between the QCRB with the photon
losses and the one without both the LDOs and the photon losses, i.e., $%
\widetilde{\Delta }=\log _{10}\Delta \phi _{L-QCRB}-\log _{10}\Delta \phi
_{QCRB}$. If the condition of $\widetilde{\Delta }<0$\ is true, then
exploitation of the LDO can effectively improve the robustness of SU(1,1)
interferometer systems. To see this point, we contourplot the difference $%
\widetilde{\Delta }$\ as a function of $\eta $\ and $\left \vert \gamma
\right \vert $, as shown in Fig. 8. it is evident that, with the increase of
$\left \vert \gamma \right \vert $, the improved region of $\widetilde{%
\Delta }<0$\ increases and more photon losses can be tolerated.

\begin{figure}[tbp]
\label{Fig9} \centering \includegraphics[width=0.82\columnwidth]{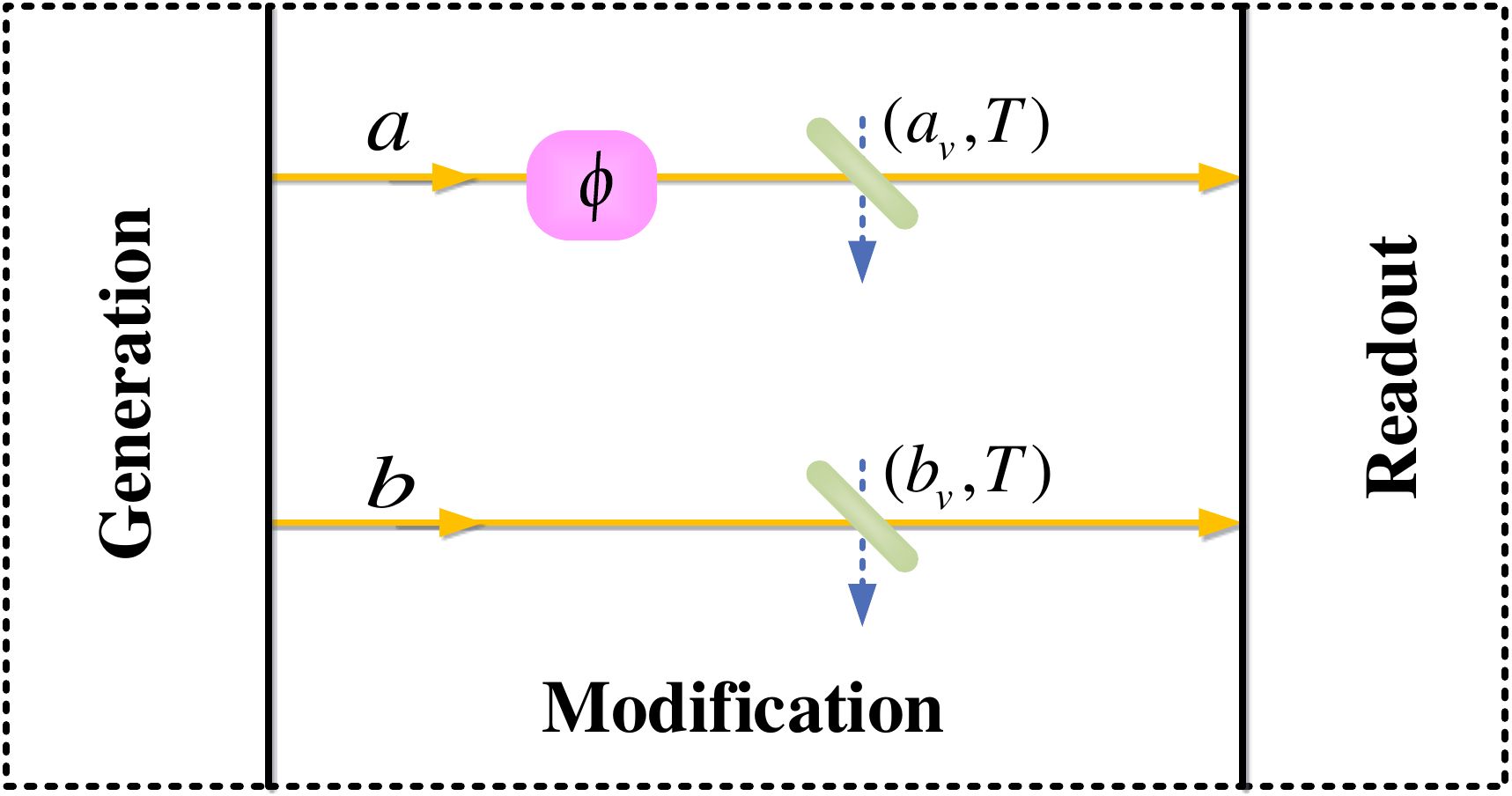}
\caption{{}(Color online) Schematic of photon-loss scenarios occurring at
after the linear phase shift in paths $a$ and $b$, in which $T$ represents
the transmissivity of the FBS. $j_{v}$ ($j=a,b$) are the vacuum operator in
path $j.$}
\end{figure}

\subsection{The effects of photon losses on the phase sensitivity}

Now, let us examine the effects of photon losses on the phase sensitivity of
DSU(1,1) interferometer. For this purpose, we assume that the same photon
losses occur at between the phase shift and the OPA$_{2}$, as shown in Fig.
9. In general, under the photon losses process, the lossy channel can be
simulated by inserting the fictitious beam splitter (FBS) with a
transmissivity $T$ \cite{17,34,53}. It is worth mentioning that, the smaller
the values of $T$, the more serious the photon losses.

For the state vector prior to the OPA$_{2}$, $\left \vert \psi _{\phi
}\right \rangle ,$ after going through the photon-loss channel, the output
state $\left \vert \psi _{\phi }\right \rangle _{out}$ in the enlarged
systems $S+E$ can be expressed as $\left \vert \psi _{out}\right \rangle =%
\hat{U}_{BS}^{a}\hat{U}_{BS}^{b}\left \vert \psi _{\phi }\right \rangle
\left \vert 0\right \rangle _{a}\left \vert 0\right \rangle _{b}$. Thus,
when passing through the OPA$_{2}$, the final output state $\left \vert \psi
_{f}\right \rangle $ can be given by%
\begin{equation}
\left \vert \psi _{f}\right \rangle =\hat{U}_{OPA2}\hat{U}_{BS}^{a}\hat{U}%
_{BS}^{b}\left \vert \psi _{\phi }\right \rangle \left \vert 00\right
\rangle _{a,b},  \label{23}
\end{equation}%
where $\left \vert 00\right \rangle _{a,b}=\left \vert 0\right \rangle
_{a}\otimes \left \vert 0\right \rangle _{b}$ is the vacuum noise, $\hat{U}%
_{OPA2}$ is the OPA$_{2}$ process, and $\hat{U}_{BS}^{\Theta }=\exp [\arccos
\sqrt{T}(\hat{\Theta}^{\dagger }\hat{\Theta}_{v}-\hat{\Theta}\hat{\Theta}%
_{v}^{\dagger })],\hat{\Theta}\in \{a,b\},$ represent the FBS operators
acting on mode $\hat{\Theta}$, with $\hat{\Theta}_{v}$ being the vacuum
noise operators. Further, by utilizing the transformations of the FBS, e.g.,
$(\hat{U}_{BS}^{\Theta })^{\dagger }\hat{\Theta}\hat{U}_{BS}^{\Theta }=\sqrt{%
T}\hat{\Theta}+\sqrt{1-T}\hat{\Theta}_{v}$, one can derive the phase
sensitivity $\Delta \phi _{L}$ with the photon losses, i.e.,

\begin{equation}
\Delta \phi _{L}=\sqrt{(\Delta \phi )^{2}+\frac{\left( 1-T\right) \cosh 2g}{%
4T(\Lambda _{1}+\Lambda _{2})^{2}}},  \label{24}
\end{equation}%
where $\Delta \phi $ can be given in the Eq. (B5) of Appendix B, and
\begin{eqnarray}
\Lambda _{1} &=&\left \vert \beta \right \vert \sinh g\cosh g\sin \left(
\phi +\theta _{\beta }\right) ,  \notag \\
\Lambda _{2} &=&\left \vert \gamma \right \vert \sinh g\sin (\phi +\theta
_{\gamma }).  \label{25}
\end{eqnarray}%
In order to explore whether the photon losses have an effect on the optimal
point $\phi $ corresponding to the minimum of phase sensitivity, we
illustrate the phase sensitivity $\Delta \phi $ with $T=0.6$ (dashed lines)
as a function of $\phi $ for several different values $\left \vert \gamma
\right \vert =0,1,2$, as shown in Fig. 10(a). As a comparison, the solid
lines represent the ideal case. As we can easily see, the minimum of phase
sensitivity is always found at the optimal point $\phi =0$, whether there is
photon loss or not. More significantly, with the increase of $\left \vert
\gamma \right \vert =0,1,2$, the gap of the phase sensitivity $\Delta \phi $
between with and without the photon losses can be further reduced around the
optimal point $\phi $. These phenomena result from that the increase of the
LDO strength $\left \vert \gamma \right \vert $ can still increase the slope
$\partial \left \langle \hat{X}\right \rangle /\partial \phi $\ of the
output signal $\left \langle \hat{X}\right \rangle $ even in the presence of
photon losses, which can be shown in Fig. 10(b).

\begin{figure}[tbp]
\label{Fig10} \centering \includegraphics[width=0.73\columnwidth]{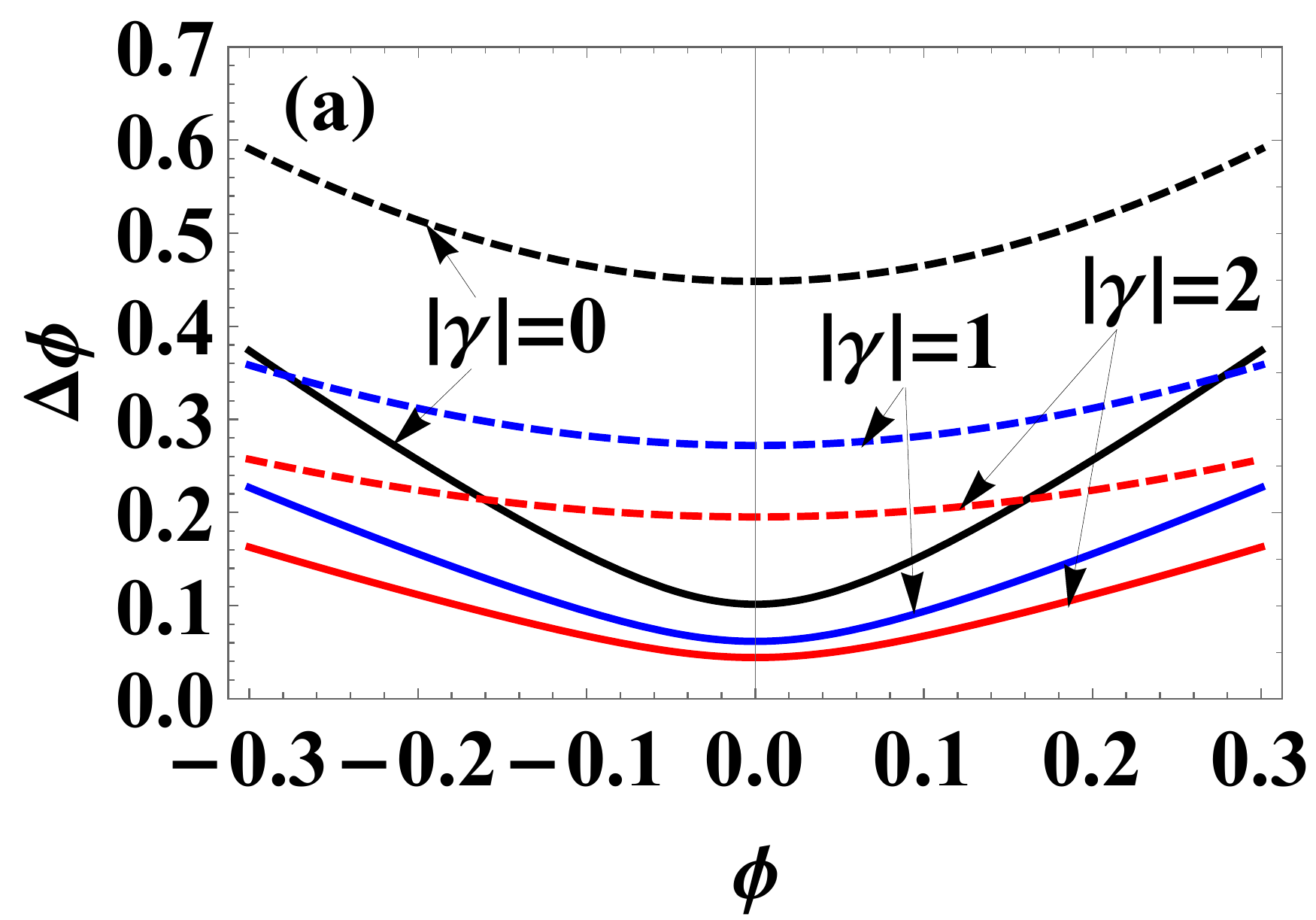}%
\newline
\includegraphics[width=0.72\columnwidth]{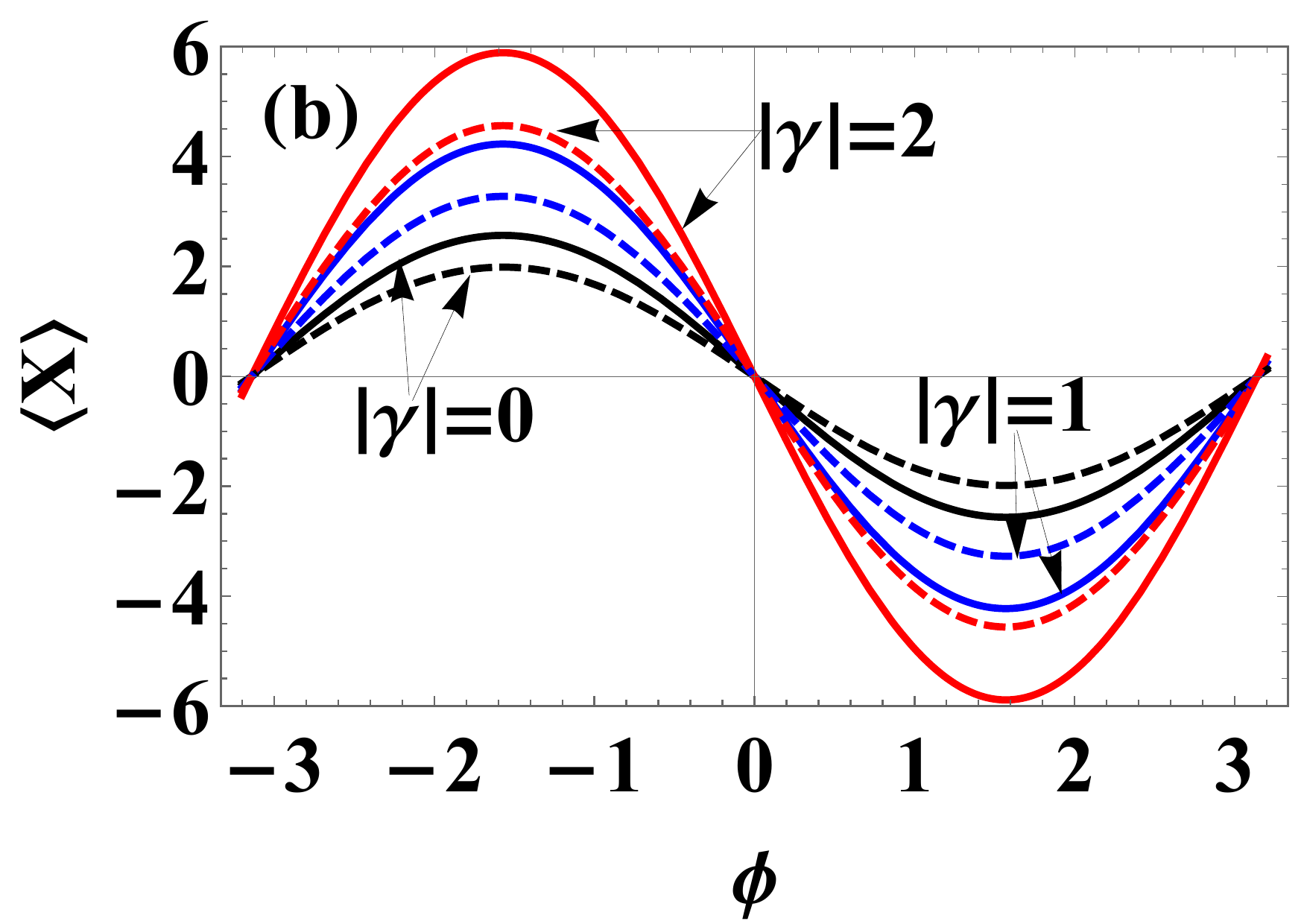}\newline
\caption{{}(Color online) When $T=0.6$, (a) phase sensitivity with homodyne
detection and (b) output signal changing with $\protect \phi $ for different $%
\left \vert \protect \gamma \right \vert =0,1,2$. As a comparison, the solid
lines are the ideal case, i.e., $T=1$. Other parameters are as following: $%
g=r=|\protect \beta |=1,$ $\protect \theta _{\protect \xi }=0$ and $\protect%
\theta _{\protect \beta }=\protect \theta _{\protect \gamma }=\protect \pi /2$.}
\end{figure}

\begin{figure}[tbp]
\label{Fig11} \centering
\subfigure[]{
\centering
\includegraphics[width=0.72\columnwidth]{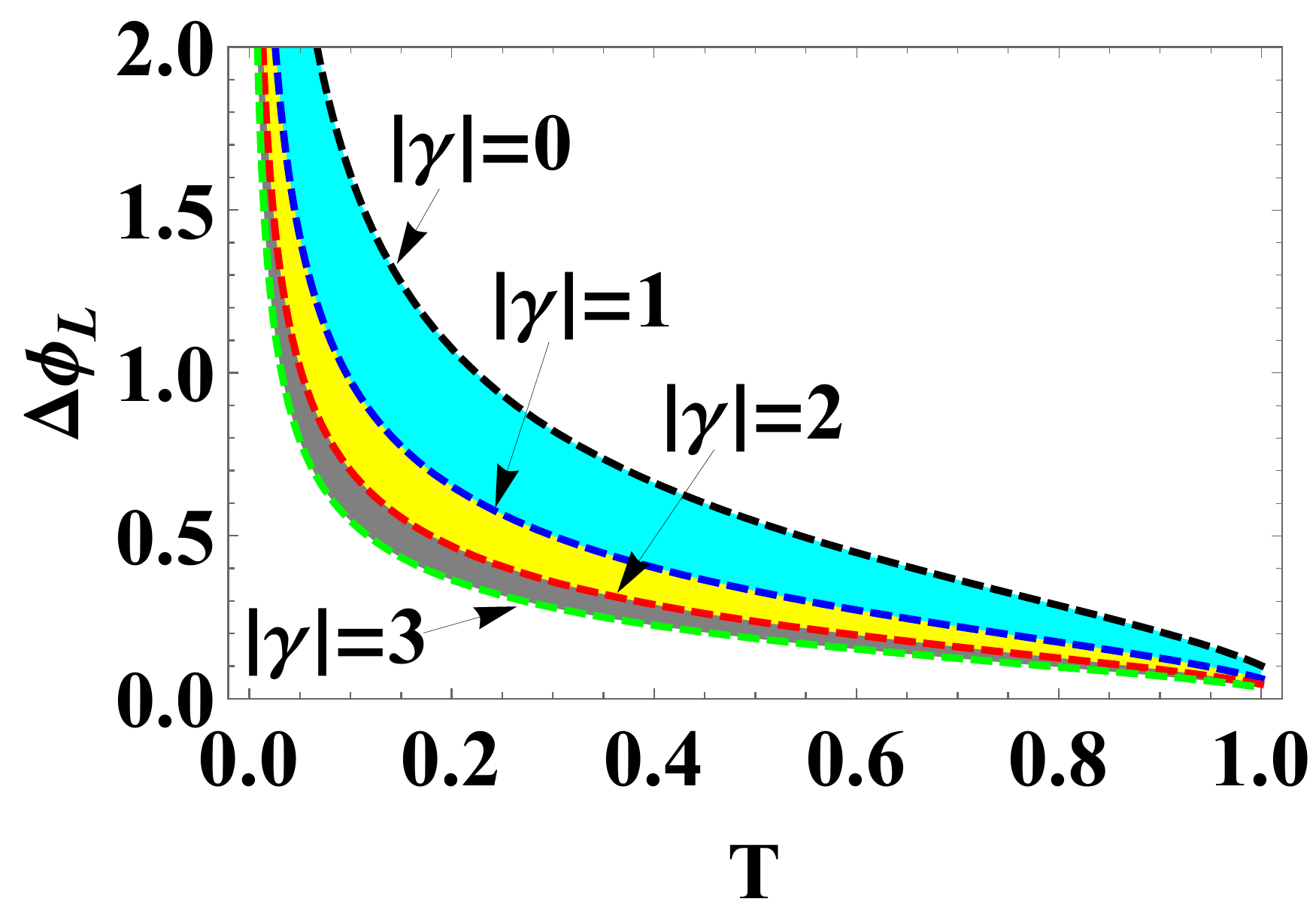}
\label{Fig11a}
}
\subfigure[]{
\includegraphics[width=0.72\columnwidth]{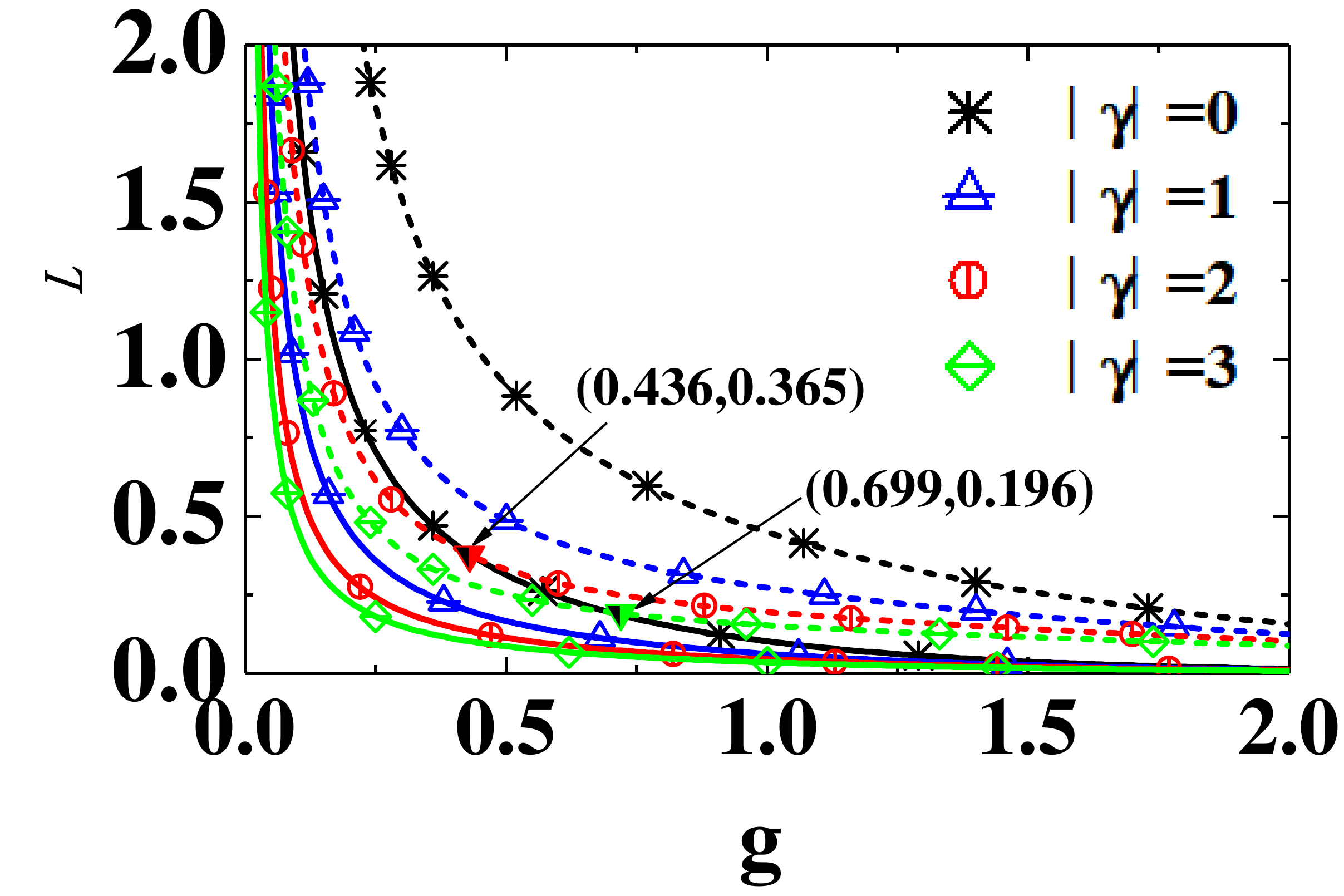}
\label{Fig11b}
}
\caption{ (Color online) Under the photon losses, the corresponding phase
sensitivity $\Delta \protect \phi _{L}$ as a function of (a) $T$ at fixed $%
\left \vert \protect \beta \right \vert =g=r=1,$ and of (b) $g$ at fixed $%
T=0.6 $, $\left \vert \protect \beta \right \vert =r=1,$ when given several
different $\left \vert \protect \gamma \right \vert =0,1,2,3$. In (b), as a
comparison, the solid lines are the ideal cases, i.e., $T=1$. Other
parameters are as following: $\protect \phi =\protect \theta _{\protect \xi }=0$
and $\protect \theta _{\protect \beta }=\protect \theta _{\protect \gamma }=%
\protect \pi /2$.}
\end{figure}
In this context, therefore, the phase sensitivity with the photon losses at
the optimal point $\phi =0$ can be calculated as%
\begin{eqnarray}
\left. \Delta \phi _{L}\right \vert _{\phi =0} &=&\left[ \frac{\left(
1-T\right) \cosh 2g}{4T(\left \vert \beta \right \vert \sinh g\cosh g+\left
\vert \gamma \right \vert \sinh g)^{2}}\right.  \notag \\
&&\left. +(\left. \Delta \phi \right \vert _{\phi =0})^{2}\right] ^{1/2},
\label{26}
\end{eqnarray}%
where the second term of the square root derives from the photon losses, and
$\left. \Delta \phi \right \vert _{\phi =0}$ is the phase sensitivity
without the photon losses given in Eq. (\ref{9}). In particular, when $T=1$,
the corresponding phase sensitivity becomes the ideal case.

In Fig. 11(a), we show the advantage of SU(1,1) interferometer robust
against the photon losses. Obviously, with the decrease of $T$, the phase
sensitivity of DSU(1,1) interferometer systems\ would fade away, but this
decline can be further slowed by increasing $\left \vert \gamma \right \vert
=0,1,2,3$. To some extent, this phenomenon reveals that the usage of the LDO
can make the whole SU(1,1) interferometer systems more robust against the
photon losses, when comparing to the case without the LDO. To visualize this
point, we make a comparison about the phase sensitivity between with (dashed
lines) and without (solid lines) photon losses, as shown in Fig. 11(b).
Surprisingly, at the same accessible parameters, the gap for the phase
sensitivity between with and without photon losses narrows down with the
increase of $\left \vert \gamma \right \vert =0,1,2,3$, even showing that
the phase sensitivity with photon losses for $\left \vert \gamma
\right
\vert =2$ $(3)$ at certain small range of $g<0.436$ ($0.699$)
performs better than that without both the photon losses and the LDO (black
solid line). In addition, the aforementioned gap can be further reduced by
increasing the value of $g.$

\section{Conclusions}

In summary, we have presented the positive contribution of the LDO for
improving the estimation performance of SU(1,1) interferometer in terms of
both the QFI and the phase sensitivity based on homodyne detection. The
numerical results show that the increase of the LDO strength is conducive to
the enhancement of the QFI and the phase sensitivity. In particular, for the
ideal case, the phase sensitivity of DSU(1,1) interferometer scheme can
gradually approach to the HL. From a realistic point of view, we further
investigate both the QFI and phase sensitivity in the presence of photon
losses. Our analyses indicate that, when given the same parameters, the
DSU(1,1) interferometer scheme can also obtain the higher QFI and the better
phase sensitivity than the SU(1,1) interferometer without the LDO under the
photon-loss case. More interestingly, the sufficiently large LDO can
strengthen the robustness of SU(1,1) interferometer systems against the
photon losses.

\begin{acknowledgments}
This work was supported by the National Natural Science Foundation of China
(Grant Nos. 91536115, 11534008, 62161029), Jiangxi Provincial Natural Science
Foundation (Grant 20202BABL202002), and Wei Ye is supported by both the
Natural Science Foundation of Jiangxi Province Youth Fund Project and the
Scientific Research Startup Foundation (Grant No. EA202204230) at Nanchang
Hangkong University.
\end{acknowledgments}

\bigskip

\textbf{Appendix\ A: The QFI for DSU(1,1) interferometer in the ideal
scenario }

For DSU(1,1) interferometer with the input state $\left \vert \psi
_{in}\right \rangle =\left \vert \xi \right \rangle _{a}\otimes \left \vert
\beta \right \rangle _{b},$ the QFI in the ideal scenario can be given by
Eq. (\ref{5}) with $\Gamma _{m}=$ $\left \langle \psi _{\gamma }\right \vert
\hat{b}^{\dagger m}\hat{b}^{m}\left \vert \psi _{\gamma }\right \rangle
,(m=1,2)$ to be calculated. For this reason, here we need to introduce the
characteristic function (CF), in which for the any probe state $\left \vert
\psi _{\gamma }\right \rangle $ of DSU(1,1) interferometer, its CF can be
expressed as%
\begin{equation}
C_{W}(\alpha _{1},\alpha _{2})=\text{Tr}[\hat{\rho}_{\gamma }\hat{D}(\alpha
_{1})\hat{D}(\alpha _{2})],  \tag{A1}
\end{equation}%
with $\hat{\rho}_{\gamma }=\left \vert \psi _{\gamma }\right \rangle
\left
\langle \psi _{\gamma }\right \vert $ being the density operator of
the probe state and $\hat{D}(\alpha _{1})=\exp (\alpha _{1}\hat{a}^{\dagger
}-\alpha _{1}^{\ast }\hat{a})$ being the displacement operator.

Thus, the average value $\Gamma _{m}$ $=$ $\left \langle \psi _{\gamma
}\right \vert \hat{b}^{\dagger m}\hat{b}^{m}\left \vert \psi _{\gamma
}\right \rangle $ can be derived as%
\begin{equation}
\Gamma _{m}=\Omega _{m}C_{N}(0,\alpha _{2}),  \tag{A2}
\end{equation}%
where $\Omega _{m}=\left. \frac{\partial ^{2m}}{\partial \alpha
_{2}^{m}\partial (-\alpha _{2}^{\ast })^{m}}...\right \vert _{\alpha
_{2}=\alpha _{2}^{\ast }=0}$ is the partial differential operator and $%
C_{N}(0,\alpha _{2})=e^{\left \vert \alpha _{2}\right \vert
^{2}/2}C_{W}(0,\alpha _{2})$ is the normal ordering form of the CF. For
DSU(1,1) interferometer with the input state $\left \vert \psi
_{in}\right
\rangle =\left \vert \xi \right \rangle _{a}\otimes \left \vert
\beta \right
\rangle _{b},$ the corresponding probe state can be given by $%
\left \vert \psi _{\gamma }\right \rangle =\hat{D}_{a}(\gamma )\hat{D}%
_{b}(\gamma )\hat{U}_{OPA1}\left \vert \psi _{in}\right \rangle $, so that
according to Eq. (A2), one can obtain%
\begin{align}
\Gamma _{m}& =\Omega _{m}\exp \left[ -\Delta _{1}\left \vert z_{2}\right
\vert ^{2}+\Delta _{2}^{\ast }z_{2}-\Delta _{2}z_{2}^{\ast }\right.  \notag
\\
& \left. -\Delta _{3}(e^{-i\theta _{\xi }}z_{2}^{\ast 2}+e^{i\theta _{\xi
}}z_{2}^{2})\right] ,  \tag{A3}
\end{align}%
where%
\begin{align}
\Delta _{1}& =\cosh ^{2}r\sinh ^{2}g,  \notag \\
\Delta _{2}& =\gamma _{2}+\beta \cosh g,  \notag \\
\Delta _{3}& =\frac{1}{4}\sinh 2r\sinh ^{2}g.  \tag{A4}
\end{align}%
Therefore, substituting Eq. (A3) into Eq. (\ref{5}), one can obtain the
explicit expression of the QFI for DSU(1,1) interferometer in the ideal case.

\textbf{Appendix\ B: Phase sensitivity via homodyne detection}

Combining Eqs. (\ref{1}) and (\ref{7}), one can derive the variance $\Delta
^{2}\hat{X}$ as
\begin{equation}
\Delta ^{2}\hat{X}\text{=}\frac{\left \vert U\right \vert ^{2}(\cosh
2r-\sinh 2r\cos \Delta )+\left \vert V\right \vert ^{2}}{2},  \tag{B1}
\end{equation}%
where%
\begin{align}
U& =\left \vert U\right \vert e^{i\theta _{U}}  \notag \\
& =\cosh ^{2}g-e^{-i\phi }\sinh ^{2}g,  \notag \\
V& =\left( 1-e^{-i\phi }\right) \sinh g\cosh g,  \notag \\
\Delta & =\theta _{\xi }+2\theta _{U},  \tag{B2}
\end{align}%
and the derivative of $\left \langle \hat{X}\right \rangle $
\begin{equation}
\frac{\partial \left \langle X\right \rangle }{\partial \phi }\text{=}-\sqrt{%
2}\left( |\beta |\cosh g\sin \Theta _{1}+\left \vert \gamma \right \vert
\sin \Theta _{2}\right) \sinh g,  \tag{B3}
\end{equation}%
with%
\begin{align}
\Theta _{1}& =\phi +\theta _{\beta },  \notag \\
\Theta _{2}& =\phi +\theta _{\gamma }.  \tag{B4}
\end{align}

Substituting Eqs. (B1) and (B3) into the error propagation formula shown in
Eq. (\ref{8}), the explicit expression of the phase sensitivity of DSU(1,1)
interferometer in the ideal case can be given by
\begin{equation}
\Delta \phi \text{=}\frac{\sqrt{\left \vert V\right \vert ^{2}+\left \vert
U\right \vert ^{2}(\cosh 2r-\sinh 2r\cos \Delta )}}{\left \vert 2\sinh
g(\left \vert \beta \right \vert \cosh g\sin \Theta _{1}+\left \vert \gamma
\right \vert \sin \Theta _{2})\right \vert }.  \tag{B5}
\end{equation}

In particular, when $\phi =\theta _{\xi }=0$, the variance $\Delta ^{2}\hat{X%
}=e^{-2r}/2.$ Moreover, by utilizing the results given in Eq. (B3) at the
optimal phase point $\phi =0,$ one can find the absolute value of the
derivative of $\left \langle \hat{X}\right \rangle $
\begin{equation}
\left \vert \partial \left \langle \hat{X}\right \rangle /\partial \phi
\right \vert \text{=}\sqrt{2}\sinh g(\left \vert \beta \right \vert \cosh
g\sin \theta _{\beta }+\left \vert \gamma \right \vert \sin \theta _{\gamma
}).  \tag{B6}
\end{equation}%
Finally, after achieving $\sin \theta _{\beta }=\sin \theta _{\gamma }=1$ by
taking $\theta _{\beta }=\theta _{\gamma }=\pi /2$, one can obtain Eq. (\ref%
{9}).

\end{document}